\documentclass[12pt]{article}

\usepackage{lmodern}
\usepackage[T2A,T1]{fontenc}
\usepackage[utf8]{inputenc}
\usepackage{geometry}
\usepackage{bm}
\usepackage{mathtools}
\usepackage{amsmath}
\usepackage{amssymb}
\usepackage{graphicx}
\usepackage{textcomp}
\usepackage[numbers,sort&compress]{natbib}
\usepackage{hyperref}
\usepackage{color}

\usepackage{amsthm}
\usepackage{mathrsfs}
\usepackage{physics}
\usepackage{marginnote}
\usepackage{bbold}
\usepackage{slashed}
\usepackage{subcaption}
\usepackage{float}
\usepackage{hyperref}
\usepackage{tcolorbox}
\usepackage[]{todonotes}
\usepackage{graphicx}
\usepackage{xcolor}
\usepackage{soul}
\usepackage{ytableau}
\usepackage{tikz}
\usepackage{tikz-feynman}

\newcommand{\barq}{q} 
\newcommand{\barl}{\ell}

\newcommand{\colStructure}{\mathcal{C}}

\newcommand{\ii}{ i} 
\newcommand{\hdelta}{\hat{\delta}} 

\usetikzlibrary{calc}
\usetikzlibrary{arrows,decorations.pathmorphing}
\tikzfeynmanset{warn luatex=false}
\renewcommand{\[}{\begin{equation}\begin{aligned}}
\renewcommand{\]}{\end{aligned}\end{equation}}
\renewcommand{\d}{\mathrm{d}}
\def\dd{\hat \d}
\def\del{\hat \delta}

\def\cut#1{{      
		\setbox\charbox=\hbox{$#1$}
		\setbox\slabox=\hbox{$|$}
		\dimen\charbox=\ht\slabox
		\advance\dimen\charbox by -\dp\slabox
		\advance\dimen\charbox by -\ht\charbox
		\advance\dimen\charbox by \dp\charbox
		\divide\dimen\charbox by 2
		\raise-\dimen\charbox\hbox to \wd\charbox{\hss$|$\hss}
		\llap{$#1$}
}}

\newbox\charbox
\newbox\slabox

\newcommand{\tree}{\begin{tikzpicture}[thick, baseline={([yshift=-\the\dimexpr\fontdimen22\textfont2\relax] current bounding box.center)}, 
decoration={markings,mark=at position 0.6 with {\arrow{Stealth}}}]
	\begin{feynman}
	\vertex (v1);
	\vertex [right = 0.3 of v1] (v2);
	\vertex [above left = .275 and .275 of v1] (o1);
	\vertex [above right = .275 and .275 of v2] (o2);
	\vertex [below left = .275 and .275 = of v1] (i1);
	\vertex [below right = .275 and .275 of v2] (i2);
	\draw (i1) -- (v1);
	\draw (v1) -- (o1);
	\draw (i2) -- (v2);
	\draw (v2) -- (o2);
	\draw (v1) -- (v2);
	\end{feynman}
	\end{tikzpicture}}

\newcommand{\boxy}{\begin{tikzpicture}[thick, baseline={([yshift=-\the\dimexpr\fontdimen22\textfont2\relax] current bounding box.center)}, decoration={markings,mark=at position 0.6 with {\arrow{Stealth}}}]
	\begin{feynman}
	\vertex (v1);
	\vertex [right = 0.25 of v1] (v2);
	\vertex [above = 0.25 of v1] (v3);
	\vertex [right = 0.25 of v3] (v4);
	\vertex [above left = 0.15 and 0.15 of v3] (o1);
	\vertex [below left = 0.15 and 0.15 of v1] (i1);
	\vertex [above right = 0.15 and 0.15 of v4] (o2);
	\vertex [below right = 0.15 and 0.15 of v2] (i2);
	\draw (i1) -- (v1);
	\draw (v1) -- (v3);
	\draw (v3) -- (o1);
	\draw (i2) -- (v2);
	\draw (v2) -- (v4);
	\draw (v4) -- (o2);
	\draw (v1) -- (v2);
	\draw (v3) -- (v4);
	\end{feynman}	
	\end{tikzpicture}}
\newcommand{\crossbox}{\begin{tikzpicture}[thick, baseline={([yshift=-\the\dimexpr\fontdimen22\textfont2\relax] current bounding box.center)}, decoration={markings,mark=at position 0.6 with {\arrow{Stealth}}}]
	\begin{feynman}
	\vertex (v1);
	\vertex [right = 0.3 of v1] (v2);
	\vertex [above = 0.3 of v1] (v3);
	\vertex [right = 0.3 of v3] (v4);
	\vertex [above left = 0.125 and 0.125 of v3] (o1);
	\vertex [below left = 0.125 and 0.125 of v1] (i1);
	\vertex [above right = 0.125 and 0.125 of v4] (o2);
	\vertex [below right = 0.125 and 0.125 of v2] (i2);
	\vertex [above right = 0.1 and 0.1 of v1] (g1);
	\vertex [below left = 0.1 and 0.1 of v4] (g2);
	\draw (i1) -- (v1);
	\draw (v1) -- (v3);
	\draw (v3) -- (o1);
	\draw (i2) -- (v2);
	\draw (v2) -- (v4);
	\draw (v4) -- (o2);
	\draw (v4) -- (g2);
	\draw (g1) -- (v1);
	\draw (v2) -- (v3);
	\end{feynman}
	\end{tikzpicture}}
\newcommand{\triR}{\begin{tikzpicture}[thick, baseline={([yshift=-\the\dimexpr\fontdimen22\textfont2\relax] current bounding box.center)}, decoration={markings,mark=at position 0.6 with {\arrow{Stealth}}}]
	\begin{feynman}
	\vertex (v1);
	\vertex [below left = 0.17 and 0.25 of v1] (v2);
	\vertex [above left = 0.17 and 0.25 of v1] (v3);
	\vertex [above right = 0.275 and 0.2 of v1] (o1);
	\vertex [below right = 0.275 and 0.2 of v1] (i1);
	\vertex [above left = 0.15 and 0.1 of v3] (o2);
	\vertex [below left = 0.15 and 0.1 of v2] (i2);
	\draw (i1) -- (v1);
	\draw (v1) -- (o1);
	\draw (i2) -- (v2);
	\draw (v2) -- (v3);
	\draw (v3) -- (o2);
	\draw (v2) -- (v1);
	\draw (v3) -- (v1);
	\end{feynman}
	\end{tikzpicture}}
\newcommand{\triL}{\begin{tikzpicture}[thick, baseline={([yshift=-\the\dimexpr\fontdimen22\textfont2\relax] current bounding box.center)}, decoration={markings,mark=at position 0.6 with {\arrow{Stealth}}}]
	\begin{feynman}
	\vertex (v1);
	\vertex [below right = 0.17 and 0.25 of v1] (v2);
	\vertex [above right = 0.17 and 0.25 of v1] (v3);
	\vertex [above left = 0.275 and 0.2 of v1] (o1);
	\vertex [below left = 0.275 and 0.2 of v1] (i1);
	\vertex [above right = 0.15 and 0.1 of v3] (o2);
	\vertex [below right = 0.15 and 0.1 of v2] (i2);
	\draw (i1) -- (v1);
	\draw (v1) -- (o1);
	\draw (i2) -- (v2);
	\draw (v2) -- (v3);
	\draw (v3) -- (o2);
	\draw (v2) -- (v1);
	\draw (v3) -- (v1);
	\end{feynman}	
	\end{tikzpicture}}
\newcommand{\nonAbL}{\begin{tikzpicture}[thick, baseline={([yshift=-\the\dimexpr\fontdimen22\textfont2\relax] current bounding box.center)}, decoration={markings,mark=at position 0.6 with {\arrow{Stealth}}}]
	\begin{feynman}
	\vertex (v1);
	\vertex [right = 0.2 of v1] (g1);
	\vertex [below right = 0.175 and 0.25 of g1] (v2);
	\vertex [above right = 0.175 and 0.25 of g1] (v3);
	\vertex [above left = 0.275 and 0.2 of v1] (o1);
	\vertex [below left = 0.275 and 0.2 of v1] (i1);
	\vertex [above right = 0.15 and 0.1 of v3] (o2);
	\vertex [below right = 0.15 and 0.1 of v2] (i2);
	\draw (i1) -- (v1);
	\draw (v1) -- (o1);
	\draw (i2) -- (v2);
	\draw (v2) -- (v3);
	\draw (v3) -- (o2);
	\draw (v1) -- (g1);
	\draw (v2) -- (g1);
	\draw (v3) -- (g1);
	\end{feynman}	
	\end{tikzpicture}}
\newcommand{\nonAbR}{\begin{tikzpicture}[thick, baseline={([yshift=-\the\dimexpr\fontdimen22\textfont2\relax] current bounding box.center)}, decoration={markings,mark=at position 0.6 with {\arrow{Stealth}}}]
	\begin{feynman}
	\vertex (v1);
	\vertex [left = 0.2 of v1] (g1);
	\vertex [below left = 0.175 and 0.25 of g1] (v2);
	\vertex [above left = 0.175 and 0.25 of g1] (v3);
	\vertex [above right = 0.275 and 0.2 of v1] (o1);
	\vertex [below right = 0.275 and 0.2 of v1] (i1);
	\vertex [above left = 0.15 and 0.1 of v3] (o2);
	\vertex [below left = 0.15 and 0.1 of v2] (i2);
	\draw (i1) -- (v1);
	\draw (v1) -- (o1);
	\draw (i2) -- (v2);
	\draw (v2) -- (v3);
	\draw (v3) -- (o2);
	\draw (v1) -- (g1);
	\draw (v2) -- (g1);
	\draw (v3) -- (g1);
	\end{feynman}	
	\end{tikzpicture}}

\newcommand{\simtree}{{\scalebox{0.5}{\tree}}} 

\hypersetup{colorlinks=true,citecolor=blue,linkcolor=black,urlcolor=black}
\numberwithin{equation}{section}
\def\Ord{{\mathcal O}}
\def\vq{\bm q}
\def\nn{\nonumber}
\def\cO{\mathcal{O}}
\def\pC{{\hat {\mathcal{C}}}}
\def\Vmom{\iffalse\widehat{V}\fi V}

\def\alm{{ \lambda }}
\def\am{{ \Lambda }}
\newcommand{\be}{\begin{equation}}
	\newcommand{\ee}{\end{equation}}

\allowdisplaybreaks
\begin{document}
	\interfootnotelinepenalty=10000
	\baselineskip=18pt
	\hfill
	
	\thispagestyle{empty}
	\vspace{1.2cm}

	\begin{center}
		{ \bf \Large 
Yang-Mills observables:\\
from KMOC to eikonal through EFT
		}

		\bigskip\vspace{1.cm}{
			{\large 
		Leonardo de la Cruz$^a$,  Andres Luna$^b$ and Trevor Scheopner$^b$
		}
		} \\[7mm]
		{\it  
$^a$Dipartamento di Fisica e Astronomia, Universit\`{a} di Bologna \\ and INFN Sezione di Bologna, via Irnerio 46, I-40126 Bologna, Italy  \\
$^b$Mani L. Bhaumik Institute for Theoretical Physics, \\[-1mm]
Department of Physics and Astronomy, UCLA, Los Angeles, CA 90095 
		}
                  \\
	\end{center}
	\bigskip
	\bigskip

	\begin{abstract} \small
 
We obtain a conservative Hamiltonian describing the interactions of two charged bodies in Yang-Mills
through $\mathcal{O}(\alpha^2)$ and to all orders in velocity. Our calculation extends a recently-introduced framework based on
scattering amplitudes and effective field theory (EFT) to consider color-charged objects. These results are checked against the direct 
integration of the observables in the Kosower-Maybee-O'Connell (KMOC) formalism.  At the order we consider
we find that the linear and color impulses in a scattering event can be concisely described in terms of the eikonal phase,
thus extending the domain of applicability of a formula originally proposed in the context of spinning particles.

\end{abstract}
	\setcounter{footnote}{0}
	
\renewcommand{\baselinestretch}{1}	
	\newpage
	\setcounter{tocdepth}{2}
	\tableofcontents
	
	\newpage

\section{Introduction}
\label{sec:intro}

The Kosower-Maybee-O'Connell (KMOC) formalism  \cite{Kosower:2018adc, Maybee:2019jus, delaCruz:2020bbn}  is a first principle approach
to extract the classical limit, understood as the limit $ \hbar \to 0$, from on-shell
scattering amplitudes. It is based on the construction of certain observables which are well-defined at the quantum and classical 
levels.  They can be defined by considering the expectation value of certain operators 
$\mathbb{O}$ evaluated at the beginning and at the end of the scattering event. Considering the two-to-two classical scattering the 
observable associated with the operator $\mathbb{O}$ is given by 

\begin{align}
\langle\Delta O \rangle=  \bra{\Psi} S^\dagger \mathbb{O} S \ket{\Psi} - \bra{\Psi}\mathbb{O} \ket{\Psi},
\end{align}	
where $S=1+\ii T$. The ``in'' states $\ket{\Psi}$ are two-particle coherent states for momenta and color, whose function is to give 
the  notion of point particles with a sharply-defined position, momenta, and color.   To make this notion precise, the restoration of
$\hbar$'s on couplings and color factors as well as the distinction between momenta $p$ and wavenumber $\bar p$ for certain particles 
play an important role. 

Employing unitarity the observables can be written as
\begin{align}
\langle \Delta O \rangle = \ii \bra{\Psi}[\mathbb O, T] \ket{\Psi}+\bra{\Psi} T^\dagger[\mathbb O,  T] \ket{\Psi}, 
\end{align}
which can be used to derive general expressions for these observables in terms of amplitudes. In this paper we will consider the 
color charge operator $\mathbb C^a_1$ and the momentum operator $\mathbb P_1^\mu$ of one of the particles, but  of course 
the other particle can  be chosen as well. The observables associated to these operators are called the color impulse $\Delta c_1^a$ and the momentum 
impulse $\Delta p_1^\mu$.
The KMOC formalism has been applied to the study of 
waveforms \cite{Cristofoli:2021vyo}, 
soft theorems
\cite{Manu:2020zxl},  radiative gravitational observables  at two-loops \cite{Herrmann:2021lqe,Herrmann:2021tct} and adapted to study
the classical limit of thermal currents \cite{delaCruz:2020cpc}.

On the other hand, the classical limit can also be described in the language of effective field theory (EFT).  This idea was pioneered in
Ref.~\cite{Neill:2013wsa}, which proposed the application of the well-established scattering-amplitudes toolkit to the  derivation of gravitational potentials.  Later,  an EFT of non-relativistic scalar fields was developed \cite{Cheung:2018wkq}, and used to translate a one-loop scattering amplitude into the $\cO(G^2)$ canonical Hamiltonian, which is equivalent to the results of Westpfahl \cite{Westpfahl:1985tsl}.  This approach was later implemented to obtain novel results at $\cO(G^3)$ order \cite{Bern:2019nnu,Bern:2019crd,Cheung:2020gyp}.

Besides making use of the KMOC formalism or non-relativistic EFTs, various approaches have been developed to extract the dynamics of compact
objects from scattering data.  These include making use of the Lippman-Schwinger equation~\cite{Cristofoli:2019neg,Bjerrum-Bohr:2019kec},  
a heavy black hole effective theory and its generalizations \cite{Damgaard:2019lfh,Aoude:2020onz,Haddad:2020tvs},  developing a 
boundary-to-bound (B2B) dictionary \cite{Kalin:2019rwq,Kalin:2019inp},  implementing a post-Minkowskian EFT \cite{Kalin:2020mvi,Kalin:2020fhe,Dlapa:2021npj} and a worldline QFT \cite{Mogull:2020sak}.
More recently the conservative binary potential at $\cO(G^4)$ was obtained by means of an amplitude-action relation that allows the calculation of 
physical observables directly from the scattering amplitude \cite{Bern:2021dqo}. 

The techniques mentioned above have been extended in multiple directions in recent years,  including the computation of observables in 
supergravity ~\cite{Caron-Huot:2018ape,Bern:2020gjj,Parra-Martinez:2020dzs} and other generalizations of GR \cite{Carrillo-Gonzalez:2021mqj,Gonzalez:2020krh}, the study of three-body dynamics \cite{Loebbert:2020aos},  
incorporating the radiation emitted by the binary into their analysis~\cite{DiVecchia:2020ymx,DiVecchia:2021ndb,Damour:2020tta,Jakobsen:2021smu,Mougiakakos:2021ckm,Bjerrum-Bohr:2021din},  and considering tidal deformations \cite{Haddad:2020que,Aoude:2020ygw,AccettulliHuber:2020dal,Kalin:2020lmz,Cheung:2020sdj,Cheung:2020gbf,Bern:2020uwk} and spin effects \cite{Holstein:2008sx,Vaidya:2014kza,Guevara:2017csg,Guevara:2018wpp,Chung:2018kqs,Chung:2019duq,Chung:2020rrz,Guevara:2019fsj,Vines:2017hyw,Bern:2020buy,Liu:2021zxr,Kosmopoulos:2021zoq,Jakobsen:2021lvp,Chen:2021huj,Bautista:2021wfy,Aoude:2021oqj,Chiodaroli:2021eug} of the astrophysical objects.

A further relation between amplitudes and classical  observables is given through the eikonal phase, which is obtained as the Fourier transform to impact parameter space of the scattering amplitude~\cite{Amati:1990xe}.  In turn,  one can derive the scattering angle through differentiation of the eikonal phase.  This subject has seen renewed interest \cite{Melville:2013qca,Luna:2016idw,Akhoury:2013yua,KoemansCollado:2019ggb,
Cristofoli:2020uzm,DiVecchia:2019myk,DiVecchia:2019kta,Bern:2020gjj,Parra-Martinez:2020dzs,DiVecchia:2021bdo,Heissenberg:2021tzo,Damgaard:2021ipf} and  a  recent calculation
in Ref.~\cite{Bern:2020buy} showed a surprising structure for the expression of the observables in terms of the eikonal phase.  This formula was 
the first example of such a relation for arbitrary orientations of the spins\footnote{Before this, there was evidence for such a relation in the
special kinematic configuration where the spins of the particles are parallel to the angular momentum of the 
system~\cite{Guevara:2018wpp,Vines:2018gqi,Siemonsen:2019dsu}.}.  This striking observation potentially implies that 
all physical observables are obtainable via simple manipulations of the scattering amplitude.   

While most of the attention has been given to gravitational theories,  Yang-Mills theory shares many important physical features with gravity,  like non-linearity and a gauge structure. Furthermore, the double copy relates scattering amplitudes in both theories\footnote{The double copy has been reviewed thoroughly in Ref.~\cite{Bern:2019prr}.}.  The connection has showed to be deeper than this, holding in a classical worldline setting \cite{Goldberger:2016iau,Goldberger:2017frp,Goldberger:2017vcg,Goldberger:2017ogt,Chester:2017vcz, Shen:2018ebu,Luna:2016hge,Luna:2017dtq,CarrilloGonzalez:2018ejf,Plefka:2018dpa,Plefka:2019hmz}, and extending to exact maps \cite{Monteiro:2014cda,Luna:2018dpt}\footnote{The classical double copy has also made contact with fluid dynamics, as shown in Refs.~\cite{Keeler:2020rcv,Cheung:2020djz}.}.  Then, since perturbation theory in Yang-Mills is far simpler than in standard approaches of gravity,  one may study Yang-Mills as a toy model for gravitational dynamics or as a building block that could be double copied to gravity.  One may also note that, as already pointed out in Ref.~\cite{delaCruz:2020bbn},  the dynamics of the color degrees of freedom in Yang-Mills, is in many respects analogous to spin (though actually simpler). This analogy with spin will be evidenced in a generalization of the formula of Ref. \cite{Bern:2020buy},  now describing the dynamics of color charges.

The proliferation of approaches to extract classical information from quantum scattering amplitudes motivates us to strive for an understanding of the relations between them.  The goal of this paper is to use Yang-Mills theory as a toy model to study the connection between three such approaches. Namely,  the KMOC formalism, the Hamiltonian approach to classical dynamics,  and a formula directly relating the eikonal phase with classical observables.

The remainder of this paper is structured as follows:
In Section \ref{sec:KMOC} we compute color and momentum impulse at NLO using the integrands obtained in Ref.~\cite{delaCruz:2020bbn}.
Then,  In Section  \ref{sec:full}, we develop the Hamiltonian approach to classical dynamics. First, we show the necessary full-theory amplitudes and use a matching procedure to an EFT to obtain the desired 
two-body Hamiltonian. Then  we use the derived Hamiltonian to compute scattering observables,  and 
check their match both to the KMOC approach of Section \ref{sec:KMOC}, as well as to the conjecture 
of Ref.~\cite{Bern:2020buy}, which directly relates these observables to the eikonal phase, 
and holds (almost unalteredly) when we include color effects.
We present our concluding remarks in Section \ref{sec:conclusions}.

\section{KMOC approach to color observables}
\label{sec:KMOC}

In this Section we introduce the KMOC approach for color and introduce our notation and conventions.
The classical scattering of two color-charged scalar particles of masses $m_1$ and $m_2$ can be modeled by
the action
\begin{align}
S=\int \dd^4 x\Big[\sum_{i=1,2} ((D_\mu \varphi_i)^\dagger (D^\mu \varphi_i)-
\frac{m^2_i}{\hbar^2} \varphi^\dagger_i \varphi_i )-\frac{1}{4} F^{a}_{\mu\nu} F^{a\mu\nu} \Big], 
\end{align}
where $D_\mu=\partial_\mu+\ii g A^{a}_\mu T^a_R$ and  $\dd^n x= (\d^n x)/(2\pi)^n$. The generators $T_a^R$ of the Lie algebra of $SU(N)$  are in some  representation $R$. 
The color charge operators, obtained from the Noether procedure, satisfy the usual Lie algebra modified by a factor of $\hbar$ 
\begin{align}
 [\mathbb C^a, \mathbb C^b]=\ii \hbar f^{abc} \mathbb C^c,
 \label{Lie-hbar}
\end{align}
emphasizing that $\mathbb{C}^a$ corresponds to an operator and 
\begin{align}
 \langle p_i|  \mathbb C^a |p^j \rangle  \equiv (C^a)_i^{\; j}=\hbar (T^a_R)_i^{\; j}.
\end{align}
So the color factors $(C^a)_i^{\; j}$ are simply rescalings of the usual generators $(T^a_R)_i^{\; j}$.  The classical color charges are then defined by
\begin{align}
c^a\equiv \bra{\psi}\mathbb C^a \ket{\psi},  
\end{align}
where the states $\ket{\psi}$ are coherent states for $SU(N)$, whose explicit  form will not be relevant for our purposes\footnote{ When considering the classical limit of multi-particle states, the full state is a tensor product of coherent
states for the kinematics and coherent states for color. $SU(N)$ coherent states can be constructed using Schwinger bosons \cite{delaCruz:2020bbn}.}. These states ensure the correct behavior of color charges in the  classical limit, namely 
\begin{align}
\bra{\psi}\mathbb C^a \ket{\psi}= &\text{finite},  \label{col-prop-1} \\
\bra{\psi}\mathbb C^a \mathbb C^b \ket{\psi}=& c^a c^b+ \text{negligible},
\label{col-prop-2}
\end{align}
which is guaranteed by choosing the dimension of the representation $R$ to be large.  The factors of $\hbar$ in Eq.\eqref{Lie-hbar} produce a nontrivial interplay
between color factors and kinematics in the classical limit. However ultimately classical quantities do not have any factors of $\hbar$ as it should be. Thus, for the 
purposes of this paper we will quote the integrands derived in Ref.~\cite{delaCruz:2020bbn} dropping the bar notation for wavenumbers. We will also employ 
the notation $ \Delta O^{(L)}$ to indicate the $L$-loop contribution to the observable such that the full result is given by
\begin{align}
 \Delta O= \Delta O^{(0)} + \Delta O^{(1)}+ \dots \,.
\end{align}
We also introduce the following notation for the Dirac-delta 
\begin{align}
 \del (x) = 2 \pi \delta (x), 
\qquad \del' (x)=\frac{ \ii }{(x-\ii \epsilon)^2}-   \frac{ \ii }{(x+\ii \epsilon)^2}. 
\end{align}
\subsection{Leading order}
Let us briefly review the LO calculation of Ref. \cite{delaCruz:2020bbn} in order to introduce some notation. We define the integral
\begin{align}
\mathcal I_{\simtree} \equiv  \int \dd^4 q \frac{\hdelta (q \cdot u_1) \hdelta (q \cdot u_2 )}{q^2} e^{-\ii q\cdot b},
\label{LO-integral}
\end{align}
where  $p_i^\mu=m_i u_i^\mu$ and $b^\mu$ is the impact parameter.  Recalling that $b^\mu$ is spacelike we also define 
$|b| \equiv \sqrt{-b^2}$. The classical four 
velocities $u_i$ are normalized to $u_i^2=1$.
The divergent integral $I_{\simtree}$ can be regulated using a
cut-off regulator $L$
\begin{align}
\mathcal  I_{\simtree}=  \frac{1}{4\pi \sqrt{\sigma^2-1} } \log \left(\frac{|b|^2}{L^2} \right),
\end{align}
where $\sigma$ is the standard Lorentz factor $\sigma=u_1 \cdot u_2$. The LO momentum impulse can then be written as 
\begin{align}
  \Delta p_1^{(0),\mu} = - g^2  \sigma c_1 \cdot c_2 \frac{\partial \mathcal  I_{\simtree}}{ \partial b_\mu},
\end{align}
where $c_1 \cdot c_2 \equiv c_1^a c_2^a$. So the momentum impulse is given by
\begin{align}
 \Delta p_1^{(0),\mu}= - 2\alpha \, c_1\cdot c_2 \frac{\sigma}{\sqrt{\sigma^2-1}} \frac{b^\mu}{b^2}, 
\end{align}
where $\alpha\equiv g^2/(4\pi)$. Similarly the color impulse at leading order reads 
\begin{align}
 \Delta c_1^{(0),a}=  g^2 \sigma  f^{abc} c_1^b c_2^c  \mathcal  I_{\simtree}=  
 \alpha\, f^{abc} c_1^b c_2^c \frac{ \sigma }{\sqrt{\sigma^2-1}} \log \left(\frac{|b|^2}{L^2} \right). \label{eq:colimpLO}
\end{align}
The divergence of the color impulse is the familiar divergence due to the long-range nature of $1/r^2$ forces in four-dimensions. 

\subsection{Next-to-Leading-Order}

The NLO momentum impulse can be obtained from the  QED one computed in  Ref.~\cite{Kosower:2018adc} using the 
charge to color replacements $Q_1Q_2 \to c_1\cdot c_2$ and $e \to g$. That this replacement works 
follows from the color-decomposition of the QCD amplitude and $\hbar$-counting as detailed in \cite{delaCruz:2020bbn}. The result reads
\[
\Delta p_1 ^{\mu,(1)} &=  \ii \frac{g^4  (c_1\cdot c_2)^2 }{2} \int \!\dd  ^4 \barl\,\dd  ^4 \barq\,   \frac{ \del (u_1\cdot \barq)\del (u_2\cdot\barq)
}{\barl^2 (\barl- \barq)^2} e^{-\ii q \cdot b}  \Bigg[ \barq^{\mu} \left\{  \frac{ \del (u_2\cdot \barl) }{m_1}  +  \frac{\del (u_1\cdot \barl)}{m_2} \right.
\\
&\qquad\qquad +  \left.  (u_1\cdot u_2)^2 \barl\cdot (\barl -\barq) \left(   \frac{\del (u_1\cdot \barl)}{m_2(u_2\cdot \barl -i\epsilon)^2} +  \frac{\del (u_2\cdot \barl) }{m_1(u_1\cdot\barl +i\epsilon)^2} \right) \right\}
\\
&\qquad\qquad -\ii  ( u_1\cdot u_2)^2 \barl ^{\mu} \barl\cdot (\barl -\barq)\left( \frac{  \del '(u_1\cdot \barl)
\del (u_2 \cdot \barl)}{m_1} - \frac{\del (u_1 \cdot \barl) \del ' (u_2 \cdot \barl)}{m_2} \right)    \Bigg ] \,.
\label{eq:colimpfull}
\]
On the other hand the NLO color impulse is given by 
\[
\Delta c_1^{a,(1)} &= g^4\!\int\! \dd ^4 \barq\, \dd  ^4 \barl\, \del (u_1\cdot \barq)\del (u_2\cdot \barq) e^{-i\barq \cdot b} \frac{1}{\barl^2 (\barl -\barq)^2}  
\\ &\quad\times\Bigg\{ \del (u_1\cdot \barl)
\Bigg[\frac{f^{acd}c_1^cc_2^d (c_1\cdot c_2)}{m_2}   \bigg[ 1+  (u_1\cdot u_2)^2 \barl\cdot (\barl - \barq) 
\bigg(\frac{1}{(u_2\cdot \barl -\ii\epsilon)^2} 
\\ 
& \qquad\qquad\qquad+ \ii \del '(u_2\cdot \barl) \bigg)\bigg]
- f^{acd}f^{dbe}c_1^b c_1^c c_2^e \frac{( u_1\cdot u_2)^2}{2} \del (u_2\cdot \barl) \Bigg]
\\ 
&\qquad+\del (u_2\cdot \barl) \Bigg[\frac{f^{acd} c_1^c c_2^d (c_1\cdot c_2)}{m_1} \bigg[ 1 + (u_1\cdot u_2)^2 \barl\cdot (\barl - \barq)\bigg( \frac{1}{(u_1\cdot \barl +\ii\epsilon)^2}  \\ 
& \qquad\qquad\qquad- \ii \del '(u_1\cdot \barl)\big)\big] + f^{acd} f^{dbe}c_1^ec_2^bc_2^c  \frac{( u_1\cdot u_2)^2}{2} \del (u_1\cdot \barl)\Bigg] \Bigg\}\,.
\label{eq:colimp}
\]

Inspecting Eqs.\eqref{eq:colimpfull} and \eqref{eq:colimp}   it is easy to see that the color and momentum impulses can be expressed in terms of the 
following ``master integrals'' 
\begin{align}
\mathcal I^{i}_{\triangle}[\alpha ,\beta ,\gamma]=&
\!\int\! \dd ^4 \barq\, \del (u_1\cdot \barq)\del (u_2\cdot \barq) e^{-i\barq \cdot b}  \int  \dd  ^4 \barl\, 
\frac{\del (u_i\cdot \barl) }{[\barl^2]^\alpha [(\barl -\barq)^2]^\beta [(\barl \cdot u_j + (-1)^i \ii \epsilon)]^\gamma}, j\ne i
\label{master-1}\\
\mathcal  I_{\slashed \Box}[\alpha,\beta]=& \!\int\! \dd ^4 \barq\, \del (u_1\cdot \barq)\del
 (u_2\cdot \barq) e^{-i\barq \cdot b} \int  \dd  ^4 \barl\, 
 \frac{\del (u_1\cdot \barl) \del (u_2\cdot \barl)}{[\barl^2]^\alpha [(\barl -\barq)^2]^\beta},
 \label{master-2}
\end{align}
where the vector dependence on the momentum transfer $q^\mu$ can be recovered by taking derivatives w.r.t. the impact parameter 
$b^\mu$. Notice that we have excluded from the master integrals those involving $\del'(x)$ since they can be reduced to the above 
cases using the identity 
\begin{align}
 \int \dd x \ x \  \del'(x) f(x^2)=- \int \dd x \ \del (x) f(x^2).
\end{align}
Following  arguments by K\"alin-Porto \cite{Kalin:2020mvi}, the integrals below vanish due to the presence 
of a double pole on a convergent integral\footnote{This result can also be shown by first using the Dirac-delta constraint and 
then IBP identities. As emphasized  by K\"alin-Porto these integrals do contribute in $d >4$ \cite{Cristofoli:2020uzm}.}
\begin{align}
\mathcal  I^{i}_{\triangle}[1,1,2]= \mathcal  I^{i}_{\triangle}[0,1,2]= \mathcal  I^{i}_{\triangle}[1,0,2]=0, \qquad i=1,2,
 \label{vanishing-ints}
\end{align} 
and therefore only $ I^{i}_{\triangle}[1,1,0]$ contributes to the observables. In the following we then simply write 
$\mathcal  I^{i}_{\triangle}[1,1,0]\equiv \mathcal  I^{i}_{\triangle}$ and for later purpose we 
write $\mathcal I_{\slashed \Box}[1,1]\equiv \mathcal I_{\slashed \Box}$.  We also have that 
\begin{align}
\mathcal  I_{\slashed \Box}[1,0]=\mathcal  I_{\slashed \Box}[0,1]=0
  \label{vanishing-ints-box}
\end{align}
since their loop integrals reduce to massless tadpole integrals. 
Now let us move on with the reductions of integrals of the form
\begin{align}
 I^\mu = \int  \dd  ^4 \barl\, 
 \barl^\mu \barl \cdot (\barl-q)
  \frac{\del' (u_1\cdot \barl) \del (u_2\cdot \barl)}{\barl^2 (\barl -\barq)^2},
  \label{vectorintegral}
\end{align}
which appear in Eq.~\eqref{eq:colimpfull} and its mirror obtained by $1 \leftrightarrow 2$.
In contrast to the above vanishing integrals, the presence of the numerator
makes this integral nonzero. Let us also recall that they are still integrated over the momentum transfer $q$ and therefore  
in the integral reduction we can set to zero any term proportional to Eq.\eqref{vanishing-ints} or \eqref{vanishing-ints-box}.
Performing a simple Passarino-Veltman reduction we can write
\begin{align}
 I^\mu = K_1 u_1^\mu+ K_2 u_2^\mu+ K_3 q^\mu,
\end{align}
where setting up a system of equations the resulting coefficients are 
\begin{align}
 K_1 = \frac{1}{1-\sigma^2} u_1 \cdot I, \qquad K_2=-\frac{\sigma}{1-\sigma^2} u_1 \cdot I, 
 \qquad K_3= \frac{1}{q^2}q\cdot I, 
\end{align}
where we have used the delta constraints $\hdelta(q \cdot u_1)$ and $\hdelta(q \cdot u_2)$  on which the
integral is supported. The result thus depends only on two integrals, namely $u_1\cdot I$ and $q\cdot I$. After cancellations, the product $q \cdot I$ leads to
\begin{align}
 q \cdot I= \frac{1}{4}\int  \dd ^4 \barl\, 
 \left[\frac{2q^2 }{\barl^2}-\frac{(q^2)^2 }{\barl^2 (\barl -\barq)^2}\right] \del' (u_1\cdot \barl) \del (u_2\cdot \barl),
\end{align}
which can be set to zero after integration over $q$ using Eq.\eqref{vanishing-ints}. Therefore we can 
express Eq.\eqref{vectorintegral} only in terms of the integral
\[
u_1\cdot I&= \int  \dd  ^4 \barl\, 
 u_1 \cdot \barl \ \ \barl \cdot (\barl-q)
  \frac{\del' (u_1\cdot \barl) \del (u_2\cdot \barl)}{\barl^2 (\barl -\barq)^2}\\
  &=-
  \int  \dd  ^4 \barl\, 
 \barl \cdot (\barl-q)
  \frac{\del (u_1\cdot \barl) \del (u_2\cdot \barl)}{\barl^2 (\barl -\barq)^2}.
\]
Without loss of generality, the second equality can be checked by choosing a frame where $u_1=(1, 0,0,0)$ and 
$u_2=(\sigma, 0,0, \sigma \beta)$ and
$\beta$ is defined from the condition $\sigma^2-\sigma^2\beta^2=1$. We can further reduce this integral ignoring vanishing terms (i.e., 
terms which have the form \eqref{vanishing-ints-box}) thus  obtaining
\begin{align}
u_1\cdot I= \frac{1}{2}  q^2   
\int  \dd  ^4 \barl\,   \frac{\del (u_1\cdot \barl) \del (u_2\cdot \barl)}{\barl^2 (\barl -\barq)^2}.
\end{align}
The result for $I^\mu$ then reads
\begin{align}
I^\mu= \frac{1}{2} q^2 \left( \frac{1}{1-\sigma^2}u_1^\mu- \frac{\sigma}{1-\sigma^2}u_2^\mu \right)\int  \dd  ^4 \barl\,   \frac{\del (u_1\cdot \barl) \del (u_2\cdot \barl)}{\barl^2 (\barl -\barq)^2},
\end{align}
which  implies that we can express our results only in terms of the integrals \eqref{master-1}-\eqref{master-2} 
as claimed. Therefore, excluding all vanishing contributions, the impulses in terms of the master integrals can be written as  
\begin{align}
\nonumber 
 \Delta p_1 ^{\mu,(1)}=  \frac{g^4 (c_1 \cdot c_2)^2}{2} \bigg\{  -\frac{\partial }{\partial b_\mu}&
 \left[\frac{ \mathcal I^{1}_{\triangle}}{m_2} +
 \frac{\mathcal  I^{2}_{\triangle}}{m_1}  \right]\\*
&- \left[\frac{\sigma^2}{2  (1-\sigma^2)}  
\left( \frac{u_1^\mu}{m_1}- \frac{\sigma u_2^\mu }{ m_1} \right)-(1\leftrightarrow 2)\right]\frac{\partial}{\partial b_\nu} 
\frac{\partial}{\partial b^\nu} \mathcal  I_{\slashed \Box}    \bigg\}
\end{align}
and 
\[
\Delta c_1^{a,(1)} &= g^4 \bigg\{ f^{acd}c_1^cc_2^d (c_1\cdot c_2) \left( \frac{
\mathcal  I_\triangle^1}{m_2}+  \frac{\mathcal  I_\triangle^2}{m_1}
\right)+
\frac{\sigma^2}{2} \left(f^{acd} f^{dbe}c_1^ec_2^bc_2^c- f^{acd}f^{dbe}c_1^b c_1^c c_2^e\right) \mathcal  I_{\slashed \Box}  
\bigg\}.
\]

Let us now consider the integration of the master integrals. The triangle one is well-known
(see e.g., Ref. \cite{Bjerrum-Bohr:2018xdl}) and we simply quote the result
\begin{align}
 \mathcal  I^1_{\triangle}=&\!\int\! \dd ^4 \barq\, \int  \dd  ^4 \barl\, 
  \del (u_1\cdot \barq)\del (u_2\cdot \barq) e^{-i\barq \cdot b} \frac{\del (u_1\cdot \barl) }{\barl^2 (\barl -\barq)^2}
  =\frac{1}{16 \pi} \frac{1}{\sigma \beta |b|} \,.
\end{align}
The loop integral inside $\mathcal  I_{\slashed \Box}[1,1]$
can be computed using dimensional regularization \cite{Bjerrum-Bohr:2021vuf}, leading to 
\begin{align}
\int  \dd  ^D \barl\,  \frac{\del (u_1\cdot \barl) \del (u_2\cdot \barl)}{\barl^2 (\barl -\barq)^2}= 
\frac{1}{2\pi \sigma \beta q^2} \left[\frac{1}{\varepsilon}- \log (-q^2)\right],
\end{align}
where $D=4-2\varepsilon$ and the usual factors $ \mu^{2\varepsilon}e^{\varepsilon\gamma_E}$ have been used to avoid the proliferation of the 
Euler-Mascheroni constant $\gamma_E$ and factors of $\pi$. The divergent term  leads to a contact term that can be discarded in the classical limit\footnote{Notice that the factor of $q^2$ in the denominator cancels after taking derivatives with respect to the impact parameter, so the singular term    leads to 
$\delta^2(\mathbf{b})$ which we can set to zero because we assume $ \mathbf b \ne 0$. }. Therefore, keeping only the 
finite part we have  
\begin{align}
\frac{\partial}{\partial b_\nu} \frac{\partial}{\partial b^\nu}  \mathcal I_{\slashed \Box} = 
 \frac{1}{2\pi \sigma \beta} \!\int\! \dd ^4 \barq\, \del (u_1\cdot \barq)\del (u_2\cdot \barq) e^{-i\barq \cdot b} 
 \log (-q^2)=\frac{1}{2 \pi^2 \sigma^2 \beta^2 } \frac{1}{b^2}.
\end{align}
It will also be convenient to use a cut-off regularization to evaluate the divergent integral 
$ \mathcal I_{\slashed \Box}$.  Exchanging the integration orders and introducing the change of variables 
$Q= -\barl+\barq $ we have
\begin{align}
\mathcal I_{\slashed \Box} =&  \int  \dd	^4 \barl\,   \frac{\del (u_1\cdot \barl) \del (u_2\cdot \barl)}{\barl^2} e^{-i\barl \cdot b}
 \int \dd ^4 Q\,  \del (u_1\cdot Q)\del (u_2\cdot Q) e^{-i Q \cdot b}
 \frac{1}{ Q ^2},
\end{align}
which leads to the product of two integrals of the form \eqref{LO-integral}. Hence the result is simply
\begin{align}
\mathcal  I_{\slashed \Box} = \mathcal I_{\simtree}^2=\frac{1}{16 \pi^2 \sigma^2 \beta^2} \log^2\left(\frac{|b|^2}{L^2}\right). 
\end{align}
For later purposes we will express the color impulse in terms of the cut-off regulated integral. Our full integrated result for the NLO momentum impulse  then reads  
\begin{align}
\nonumber \Delta p_1 ^{\mu,(1)}=  (c_1 \cdot c_2)^2 \frac{ 2\pi \alpha^2 }{m_1 m_2} \bigg\{ -&\frac{1}{4\sqrt{\sigma^2-1}} (m_1+m_2) 
\frac{b^\mu}{|b|^3}\\ 
&-\frac{1}{\pi}\frac{1}{b^2} \frac{\sigma^2}{(\sigma^2-1)^2}
\left[(m_2 + \sigma m_1) u_1^\mu-(m_1+\sigma m_2) u_2^\mu \right]   \bigg\},
\end{align}
and for the NLO color impulse 
\begin{align}
\nonumber   \Delta c_1^{a,(1)} = \alpha^2  \bigg\{\pi&\frac{f^{acd}c_1^cc_2^d (c_1\cdot c_2)}{ \sqrt{\sigma^2-1}|b|} \Big( \frac{1}{ m_1 }+ 
   \frac{1}{ m_2} \Big)\\&+  \frac{1}{2} \frac{ \sigma ^2 }{ (\sigma^2-1)} \log^2\left(\frac{|b|^2}{L^2}\right)\Big[
f^{acd} f^{dbe}c_1^ec_2^bc_2^c-f^{acd}f^{dbe}c_1^b c_1^c c_2^e\Big]  \bigg\}. \label{eq:colimpNLO}
\end{align}

\section{Hamiltonian approach to color dynamics}
\label{sec:full}

In this Section we will compute the position-space Hamiltonian  $H$ that describes the classical dynamics of the two-to-two scattering of 
$SU(N)$ colored objects with masses $m_1$ and $m_2$ and color charges $c_1$ and $c_2$. The classical dynamics described by such a Hamiltonian must 
be consistent with Wong's 
equations \cite{Wong:1970fu} and its perturbative solutions and by extension to observables in the KMOC formalism.
Let $\bm r$ and $\bm p$  be the relative distance between the particles and   
the momentum vector in the center of mass frame, respectively. We are  interested in a perturbative expansion of the Hamiltonian
\begin{align}
H \equiv H({\bm
  r}, {\bm p}, \mathcal C_i )  = \sqrt{\bm p ^2 + m_1^2}+\sqrt{\bm p^2 + m_2^2}+V(\bm r^2, \bm p^2, \mathcal{C}_i)
+ \dots \,,
\label{Hamiltonian_general}
\end{align}
where the potential is an expansion up to the second power in the coupling constant $\alpha$ and 
the color structures $\mathcal{C}_i$  are all possible functions of the color charges that can appear in the amplitude. 
These charges  are understood in the sense of Wong, i.e., as the classical limit of a quantum operator in a
large representation of the gauge group so they can be treated as $c$-numbers.

\subsection{Classical perturbation theory}

Consider the general problem of an arbitrary Hamiltonian $H$ describing
the interaction of two particles with color charges
$c_1$ and $c_2$ in their center of mass frame.  
While, as usual, $\bm r$ and $\bm p$ are canonically-conjugate to each
other,  color charges do not have a natural canonical
conjugate. To derive the equations of motion we use the fact that they
satisfy the relation \cite{Balachandran:1976ya,Balachandran:1977ub}
\begin{equation}
\{c^a_i,\, c^b_j\}=\delta_{ij} \, f^{abc} c^c_i\,, \hskip 1.5 cm i,j=1,2 \,,
\label{PBspin}
\end{equation}
where $\{A,B\}$ is the Poisson bracket of $A$ and $B$.  
The equations of motion are then
\begin{equation}
\dot {\bm r} = \frac{\partial H}{\partial {\bm p}} \,,
\hskip 1.5 cm 
\dot {\bm p} = -\frac{\partial H}{\partial {\bm r}} \,,
\hskip 1.5 cm 
\dot c_i^a =f^{abc}c_i^b \frac{\partial H}{\partial c_i^c} ~,~~~i=1,2 \, .
\label{general_eom-2}
\end{equation}
In the color equation of motion,  no summation over $i$ is implied on the right-hand side.
For the purpose of 
finding the impulse $\Delta \bm p$ we find it convenient to use Cartesian coordinates.
One can  solve the equations of motion for coordinates, momenta, and colors 
as a function of time. 

There are conservation laws that aid the construction of classical
solutions. These fix the energy and the total angular momentum in
terms of their asymptotic values. For example for the energy we have
\begin{align}
E & \equiv  H({\bm r} _\infty, {\bm p}_\infty, c_1, c_2)=
\sqrt{\bm p_\infty^2+m_1^2}+\sqrt{\bm  p_\infty^2+m_2^2} \,,
\label{conservation}
\end{align}
where ${\bm p}_\infty = p_\infty {\bm e}_z$ is the incoming momentum
at infinity.  We take the orbital angular momentum at infinity to be
\begin{equation}
{\bm L} \equiv  {\bm b}\times {\bm p}_\infty  = b \cdot p_\infty {\bm e}_y \, ,
\end{equation}
where $\bm b = -b\bm e_x$ and $b$ is the impact parameter. 
We solve the equations of motion perturbatively in the coupling constant, i.e. we search for a 
solution for coordinates, momenta, and colors of the form
\begin{align}
\bm r(t) &= \bm r_0(t) + \alpha \bm r_1(t)+ \alpha^2 \bm r_2(t)+\dots \ , 
\nonumber\\
\bm p(t) &= \bm p_0(t) + \alpha \bm p_1(t)+ \alpha^2 \bm p_2(t)+\dots \ , \label{iteration-EOM}
\\
c^a_i(t) &= c^a_{i,0}(t) + \alpha c^a_{i,1}(t)+ \alpha^2 c^a_{i,2}(t)+\dots \ .
\nonumber
\end{align}
Replacing them in the equations of motion \eqref{general_eom-2} leads to iterative relations
between the time derivative of the $n$-th term in the expansions above and all the 
lower-order terms. The $\Ord(\alpha^0)$ terms describe the motion of a free color-charged particle in flat space,
i.e. a straight line fixed by the initial momentum, the impact parameter, and initial color charge.
The first-order differential equations for the higher-order terms can be integrated; the relevant 
boundary conditions are that $\bm r_{n\ge 1}$, $\bm p_{n\ge 1}$ and $ c^a_{i,n\ge 1}$ vanish 
at $t=-T$, where $T$ is a cutoff time. It is necessary to introduce such a cutoff due to the same divergence 
identified in Eqs.\eqref{eq:colimpLO} and \eqref{eq:colimpNLO}; the cutoff $T$ is proportional to the cutoff $L$ in those equations. The 
contribution of each order in $\alpha$ to an observable $O$, such as the linear or color impulse, is then
\begin{equation}
\Delta O^{(n)} =   \int_{-T}^T \mathrm{d}t\,\frac{\mathrm{d}O^{(n)}}{\mathrm{d}t} = O^{(n)}(t=T) - O^{(n)}(t=-T) \, ,
\label{observables-2}
\end{equation}
with the complete result being their sum weighted with the appropriate powers of $\alpha$. 

\subsection{Hamiltonian from effective field theory }
\begin{figure}
\begin{center}
\includegraphics[scale=.7]{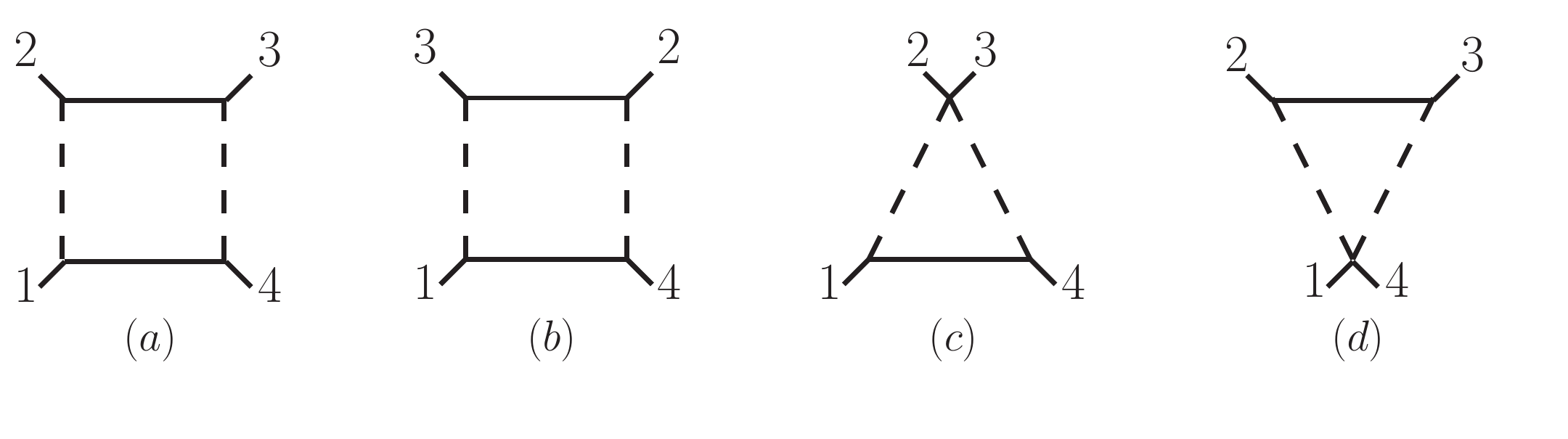}
\end{center}
\vskip -.5 cm
\caption{\small The one-loop scalar box integrals $I_{\Box}$ (a) and $I_{\bowtie}$ (b) and the corresponding triangle integrals $I_{\bigtriangleup}$ (c) and $I_{\bigtriangledown}$ (d). The bottom (top) solid line corresponds to a massive propagator of mass $m_1$ ($m_2$). The dashed lines denote massless propagators.}
\label{OneLoopIntegralsFigure}
\end{figure}

The perturbative classical problem can be solved straightforwardly once the Hamiltonian is obtained. We then proceed to compute it 
following the EFT approach adapted to this case. In order to apply this approach we will decompose the amplitudes in some color basis 
and neglect contributions of higher orders in $\hbar$ using Eq.~\eqref{Lie-hbar}. Our amplitude expressions will be directly written in 
terms of classical color factors, i.e., we consider that the expectation value with respect to coherent states has already been taken\footnote{This essentially amounts to the replacement $C_i \to c_i$ which is implemented in Ref.\cite{delaCruz:2020bbn}
by the double bracket notation.}.

\subsubsection{Full theory amplitudes from unitarity}
Let us first show the two-to-two scattering amplitudes between color-charged particles needed to construct the Hamiltonian.
The information to determine the $\mathcal{O}(\alpha)$ Hamiltonian is contained in the tree-level amplitude. We take the incoming momenta of 
the color-charged 
particles to be $-p_1$ and $-p_2$ and their outgoing momenta to be $p_3$ and $p_4$. 
The amplitude is given by 
\begin{align}
\label{treeS1S2}
\mathcal{A}^{\textrm{tree}} = 
& -\frac{4\pi \alpha}{q^2} \,
\alm_1 \mathcal{C}\left(\tree\right)
+ \ldots\,,
\end{align}
where we omit terms that do not contribute to the classical limit in the ellipsis,  along with pieces proportional to $q^2$, since they cancel
the propagator and do not yield long-range contributions. 
The color structure is given by
\begin{equation}
\colStructure\left(\tree\right) = c_1\cdot c_2\,,
\end{equation}
and the coefficient $\alm_1$ takes the explicit form
\begin{align}
\alm_1=-4m_1m_2\sigma
\,,
\label{eq:treeAlphas-2}
\end{align}
where we use the kinematic variable
\begin{equation}
\sigma= \frac{p_1\cdot p_2}{m_1m_2}\,.
\label{eq:varDefs1}
\end{equation}

In order to construct the $\mathcal{O}(\alpha^2)$ Hamiltonian we further need the corresponding one-loop amplitude. 
It was shown in Ref. \cite{delaCruz:2020bbn} that classically, the 1-loop scalar YM amplitude has a basis of only one color factor, and moreover depends on the same topologies as in electrodynamics, so it's given by
\[
\mathcal{A}^{\textrm{1-loop}} &=  \colStructure\!\left(\boxy \right) \mathcal{A}^{\textrm{1-loop,\,QED}} + \ldots\,,
\]
in terms of the one-loop QED amplitude.  The color structure is given by 
\begin{equation}
\colStructure\!\left(\boxy \right) = \left(c_1 \cdot c_2 \right)^2 \,  .
\end{equation}
We could express the latter one-loop amplitude as a linear combination of scalar box, triangle,  bubble and tadpole integrals, but
Refs.~\cite{Cheung:2018wkq,Bern:2019crd} showed that the bubble and tadpole integrals do not contribute to the classical limit.  Dropping these pieces we write
\begin{equation}
i\mathcal{A}^\text{1-loop, \,QED} = d_{\Box}\, I_{\Box} + d_{\bowtie}\, I_{\bowtie}
+ c_{\bigtriangleup} \, I_{\bigtriangleup} + c_{\bigtriangledown}\,  I_{\bigtriangledown} \, ,
\label{Org_Scalar_Integrals}
\end{equation}
where the coefficients $d_{\Box}$, $d_{\bowtie}$,
$c_{\bigtriangleup} $ and $c_{\bigtriangledown} $ are rational
functions of external momenta.  The integrals  $I_{\Box}$, $I_{\bowtie}$,
$I_{\bigtriangleup} $ and $I_{\bigtriangledown} $ are shown in 
Fig.~\ref{OneLoopIntegralsFigure}. The triangle integrals take the form~\cite{Cheung:2018wkq}

\begin{eqnarray}
I_{\bigtriangleup,\bigtriangledown}= -\frac{i}{32m_{1,2}} \frac{1}{\sqrt{-q^2}} + \cdots\,.
\label{eq:triangleIntegrals}
\end{eqnarray}
The box contributions do not contain any novel $\mathcal{O}(\alpha^2)$ information. They correspond to infrared-divergent pieces that cancel out when we equate the full-theory and EFT amplitudes~\cite{Cheung:2018wkq,Bern:2019crd}.  In this sense, the explicit values for the box coefficients serve only as a consistency check of our calculation and we do not show them. Instead, we give the result for 
\begin{align}
i \mathcal{A}^{\bigtriangleup+\bigtriangledown}\equiv  \left(c_{\bigtriangleup} \, I_{\bigtriangleup} + c_{\bigtriangledown}\,  I_{\bigtriangledown}\right)\mathcal{C}\left(\boxy\right)\, .
\label{Full_Amp}
\end{align}

As detailed in Ref.\cite{Bern:2020buy},  we use the generalized-unitarity method to obtain the integral coefficients of 
Eq.~\eqref{Org_Scalar_Integrals}.  We start by calculating the Compton amplitude for the
color-charged particle, using Feynman rules. Subsequently, we construct the two-particle cut. The residues of the two-particle cut on the 
matter poles give the triple cuts, and localizing both matter poles gives the quadruple cut.  We obtain the triangle and box coefficients from
the triple and quadruple cuts respectively.  Our result reads

\begin{figure}[tb]
\begin{center}
  \includegraphics[scale=.5]{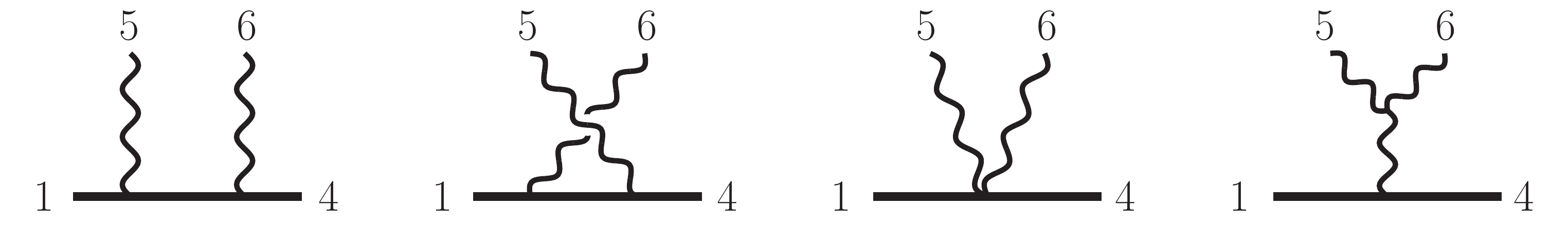}
\end{center}
\vskip -.4 cm
\caption{\small The Compton-amplitude Feynman diagrams. The straight line corresponds to the 
massive color-charged  particle. The wiggly lines correspond to gluons.  }
\label{fig:ComptonFeynRules}
\end{figure}

\begin{figure}[tb]
\begin{center}
  \includegraphics[scale=.49]{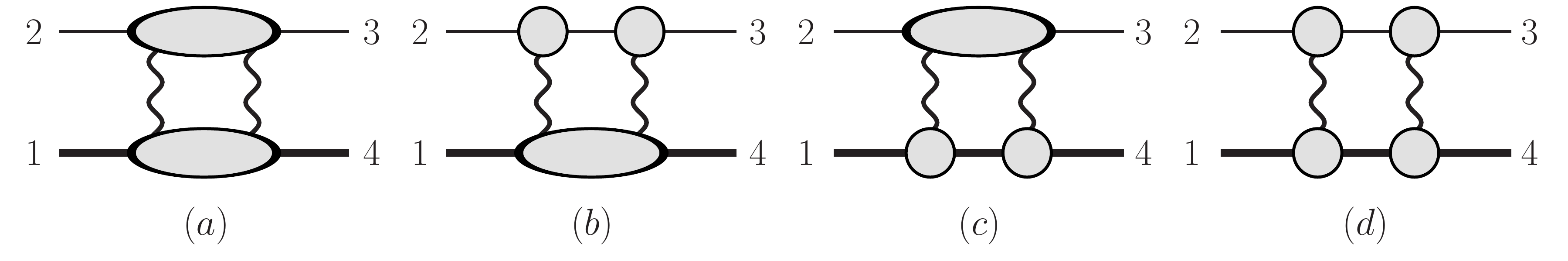}
  \put(-95,50){5} \put(-43,50){6}
  \put(-220,50){5} \put(-168,50){6}
  \put(-345,50){5} \put(-293,50){6}
  \put(-470,50){5} \put(-418,50){6}
\end{center}
\vskip -.4 cm
\caption{\small Appropriate residues of the two-particle cut (a) give the triple cuts (b) and (c), and the quadruple cut (d). The 
straight lines corresponds to the color-charged  particles and the wiggly lines to the exchanged gluons. 
All exposed lines are taken on-shell. }
\label{Cuts}
\end{figure}
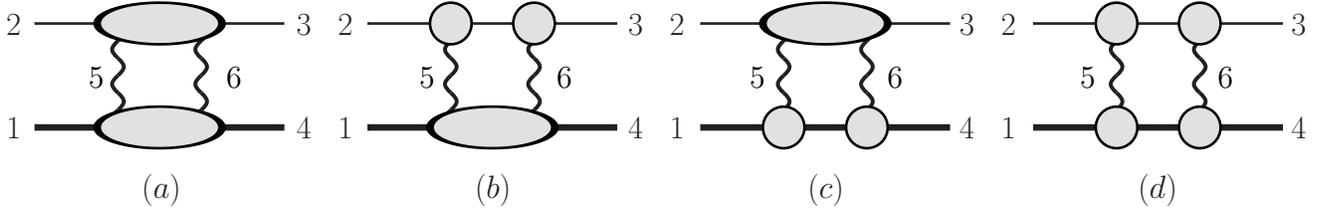

\begin{align}
\mathcal{A}^{\bigtriangleup+\bigtriangledown}
 =\frac{2\pi^2\alpha^2 }{\sqrt{-q^2}}
\alm_2 \mathcal{C}\left(\boxy\right)+\ldots\,,
\end{align}
where the coefficient is given by 
\begin{align}
\alm_2=2m
\,,
\label{eq:1loopAlphas-2}
\end{align}
and  $m=m_1+m_2$. In preparation for the matching procedure in the following Section, we specialize our expressions to the
center-of-mass frame.  In this frame, the independent four-momenta read
\begin{align}
p_1=-(E_1, \bm p)
\, ,\hskip 1.5 cm 
p_2=-(E_2, -\bm p)
\, , \hskip 1.5 cm 
q=(0, \bm q)
\, , \hskip 1.5 cm 
\bm p\cdot \bm q = \bm q^2/2\,.
\label{COMdef}
\end{align}
Using the above expressions, our amplitudes take the form
\begin{align}
\label{fullTheoryM}
\frac{\mathcal{A}^\text{tree}}{4E_1E_2}&=
\frac{4 \pi  \alpha }{\bm q^2}
\am_{1} \mathcal{C}\left(\tree\right)
\, ,\qquad
\frac{\mathcal{A}^{\bigtriangleup+\bigtriangledown}}{4E_1E_2}=
\frac{2\pi^2  \alpha^2 }{|\bm q |}
\am_{2} \mathcal{C}\left(\boxy\right) \, .
\end{align}
The coefficients $\am_i$ are given in terms of the $\alm_i$ of Eqs.~\eqref{eq:treeAlphas-2} and \eqref{eq:1loopAlphas-2} by
\begin{align}
\am_1 & =  
-\frac{\nu\sigma}{\gamma^2\xi},\qquad
\am_2 =\frac{1}{2m\gamma^2\xi}
\, ,
\label{eq:a_covRelations-2}
\end{align}
where in addition to the definition in Eq.~\eqref{eq:varDefs1} we use
\begin{equation}
\nu=\frac{m_1m_2}{m^2}\, \qquad
\gamma = \frac{E}{m} \,,\qquad
E = E_1 + E_2 \,, \qquad
\xi = \frac{E_1 E_2}{E^2} \,.
\label{eq:varDefs2}
\end{equation}

\subsubsection{Construction of the EFT amplitudes}
With the full theory amplitudes in hand,  we now turn our attention to the task of translating the scattering amplitudes of color-charged fields to a 
two-body conservative Hamiltonian. We do this by matching the scattering amplitude computed above to the two-to-two amplitude of an EFT of the positive-energy 
modes of fields.  Ref.~\cite{Cheung:2018wkq} developed this matching procedure for higher orders in the coupling constants and all orders in velocity, and we 
adapt it here to describe the color-charged fields $\xi_{1}$ and $\xi_2$. We follow closely the construction for classical spin
in Ref.~\cite{Bern:2020buy}. The action of the effective field theory (supressing representation indices) for $\xi_{1}$ and $\xi_2$ is given
by
\begin{align}
	S =& \int \dd^{D-1}{\bm k}   \, 
\sum_{a=1,2} \xi_a^\dagger(-\bm k) \left(i\partial_t - \sqrt{\bm k^2 + m_a^2}\right) \xi_a(\bm k)
 \label{eq:eftL} \\*
&-\int\dd^{D-1}{\bm k} \int \dd^{D-1}{\bm k'} \, 
 \xi_1^\dagger(\bm k') \xi_2^\dagger(-\bm k') \,
  \Vmom(\bm k' ,\bm k, \pC_i) \,\xi_1(\bm k)\xi_2(-\bm k) \, ,
\nonumber
\end{align}
where the interaction potential $\Vmom(\bm k' ,\bm k, \pC_i)$ is a function  of the incoming and outgoing momenta $\bm k$ and $\bm k'$ and the 
color-structure operators $\pC_i$.  
We consider kinematics in the center-of-mass frame. As on the full theory side, one could construct the color asymptotic states of $\xi_i$ 
using $SU(N)$ coherent states (analogous to the spin coherents states of \cite{Bern:2020buy})  so color operators   
satisy the defining properties eqs.~\eqref{col-prop-1}-\eqref{col-prop-2}. 
We obtain the classical color charge vector as the expectation value of the color operator with respect to these on-shell states.  

We build the most general potential containing only long-range classical contributions.  This will be in terms of color operators, whose
expectation values with respect to $SU(N)$ coherent states are in correspondence with the classical color structures in the full theory amplitude,
Eq.~\eqref{fullTheoryM}.  We use the following ansatz for the potential operator
\begin{align}
\hat{\Vmom}(\bm k', \bm k, \pC_i) =
& \frac{4\pi \alpha}{\hat{ \bm q}^2} d_1\left(\hat{\bm p}^2 \right)\pC\left(\tree\right)
+ \frac{2\pi^2 \alpha^2}{|\hat{\bm q}|}  d_2\left(\hat{\bm p}^2 \right)\pC\left(\boxy\right)+  \Ord(\alpha^3) \,,
\label{eq:V_mom}
\end{align}
where $\hat{\bm p}^2 \equiv ( \bm k^2 + \bm k^{\prime 2} )/2$.

We now evaluate the EFT two-to-two scattering amplitude.  To this end we use the Feynman rules derived from the EFT action 
(Eq.~\eqref{eq:eftL}),
\begin{equation}
	\includegraphics[scale=.55,trim={0 0.2cm 0 0}, clip]{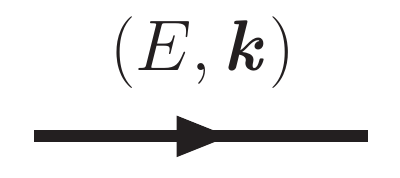} = \frac{i\, \mathbb{I}}{E-\sqrt{\bm k^2+m^2}+i\epsilon}\,, \qquad
	\vcenter{\hbox{\includegraphics[scale=.55,trim={0 0.5cm 1cm 0}, clip]{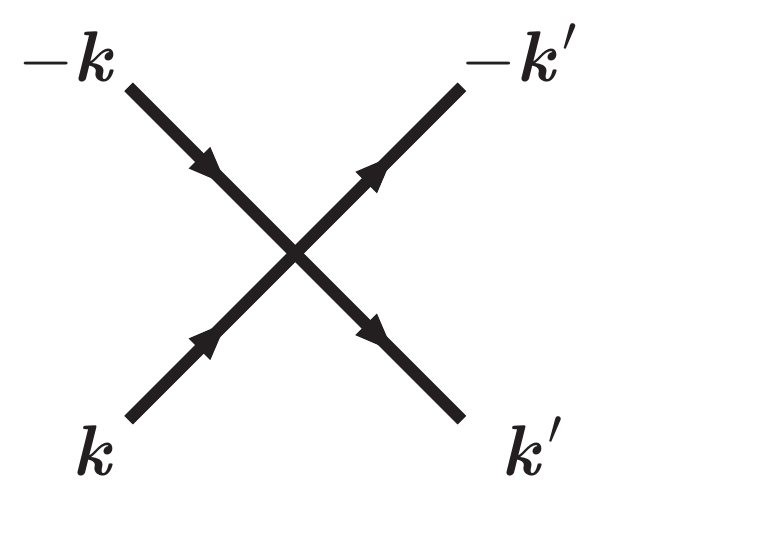}}} =
	-i V(\bm k', \bm k, \pC_{i})\,.
\label{eq:EFT_FeynmanRules}
\end{equation}

Using these rules we compute the amplitude up to $\cO(\alpha^2)$ directly evaluating the relevant Feynman diagrams, omitting terms that do not contribute to long range interactions. The color factors must be treated as operators,  and thus their ordering is important.  After 
carrying out the energy integration,  we obtain an expression  for the amplitude 
\begin{align}
\hat{\mathcal{A}}^{\textrm{EFT}}
=& - \hat{\Vmom}(\bm p',\bm p, \pC_i) - \int \dd^{D-1}{\bm k} \frac{\hat{\Vmom}(\bm p',\bm k, \pC_i)\, \hat{\Vmom}(\bm k,\bm p, \pC_i)}{E_1+E_2-
\sqrt{\bm k^2+m_1^2}- \sqrt{\bm k^2+m_2^2}} \ .
\label{eq:eft_FeynmanDiagrama}
\end{align}
We can now take the expectation value with respect to coherent states.  At $\cO(\alpha)$ the EFT amplitude receives a contribution only from the first 
term of Eq.~\eqref{eq:eft_FeynmanDiagrama}, and  after taking the  expectation value with respect to coherent states 
the result is 
\begin{align}
\mathcal{A}^{\rm EFT}_{\mathcal{O}(\alpha)} =
-\frac{4\pi \alpha }{\bm q^2}
d_1 \mathcal{C}\left(\tree\right)\,,
\label{eq:M_1PM}
\end{align}
which is a $c$-number.
On the other hand, the EFT amplitude at $\mathcal{O}(\alpha^2)$ receives contributions
from both terms in Eq.~\eqref{eq:eft_FeynmanDiagrama} and can be
written as
\begin{align}
\mathcal{A}^{\rm EFT}_{\mathcal{O}(\alpha^2)} & =
\frac{2\pi^2 \alpha^2}{|\bm q|}
\am_2\, \mathcal{C}\left(\boxy\right)
+(4\pi \alpha)^2\, \am_{\rm iter}\,\mathcal{C}\left(\tree\right)^2\int \dd^{D-1} \bm \ell
\frac{2\xi E}{\bm \ell^2 (\bm \ell+\bm q)^2 (\bm \ell^2+2\bm p\cdot \bm \ell)} \,,
\label{eq:M_2PM}
\end{align}
where $\bm \ell = \bm k - \bm p$ and we only keep terms that are relevant in the 
classical limit.
Anticipating the matching, we write the amplitude in terms of  $\am_2$, which is given directly in terms of the momentum-space potential coefficient by
\begin{align}
\label{eq:a2}
\am_{2} &=
-d_2+\frac{1-3\xi}{2\xi E}\, d_1^2  + \xi E\partial_{\bm p^2} d_1^2
 \,,
\end{align}
The second term in Eq.~\eqref{eq:M_2PM} is infrared divergent and we have explicitly verified that it cancels out when we equate the full-theory and EFT amplitudes.  The potential takes the form 
\begin{align}
V(\bm r^2, \bm p^2,\mathcal{C}_i)&=\frac{\alpha}{|\bm r|} d_1(\bm p^2)\mathcal{C}\left(\tree\right)+ \left(\frac{\alpha}{|\bm r|} \right)^2 d_2(\bm p^2)\mathcal{C}\left(\boxy\right)+ {\cal O}(\alpha^3) \,.
\end{align}
We obtain the position-space Hamiltonian by taking the Fourier transform of the momentum-space\footnote{The position-space coefficients are trivially related to the momentum-space coefficients. This is unlike the case for spinning particles, where a set of linear relations was established between them. }  Hamiltonian with respect to the momentum transfer $\vq$, which is the conjugate of the separation between the particles $\bm r$. 
We determine the momentum-space coefficient $d_i$ in terms of the amplitudes coefficients $\am_i$ by a matching procedure, i.e.  by demanding that the EFT amplitude matches the full-theory one,
\begin{align}
{\cal A}^\text{EFT}_{\mathcal{O}(\alpha)} = \frac{{\cal A}^\text{tree}}{4E_1 E_2}
\ , \qquad
{\cal A}^\text{EFT}_{\mathcal{O}(\alpha^2)}= \frac{{\cal A}^\text{1-loop}}{4E_1 E_2}
\ ,
\label{IDamplitudes}
\end{align}
where the factors of the energy account for the non-relativistic normalization of the EFT amplitude. 
Using Eq.  \eqref{eq:a_covRelations-2} we relate $\am_i$ to $\alm_i$, which are explicitly shown in Eqs.  \eqref{eq:treeAlphas-2} and 
\eqref{eq:1loopAlphas-2}.
Putting everything together, we obtain expressions for the position-space coefficients
\begin{align}
d_1&=-\frac{\nu\sigma}{\gamma^2 \xi},\\ d_2&=\frac{1}{m\xi}
\left(
\frac{1}{2\gamma^2}-
\frac{\nu\sigma}{\xi\gamma^3}+
\frac{(1-\xi)\nu^2\sigma^2}{2\xi^2\gamma^5}
\right).
\end{align} 
This finishes the computation of the effective Hamiltonian. The  classical equations of motion 
can now be solved iteratively using the Eqs.\eqref{general_eom-2}, 
\eqref{iteration-EOM} and the defintion of the observables \eqref{observables-2}. Following this procedure we have 
found agreement with the results of Section \ref{sec:KMOC}.

\subsection{Observables from the eikonal phase}
The conservative Hamiltonian we obtained in previous  Sections
enables the calculation of physical observables for a scattering of compact objects interacting through gluon exchange.  Ref.~\cite{Bern:2020buy} conjectured a
formula that expresses physical observables in terms of derivatives of the eikonal phase for the spinning case.  
In this Section we extend that analysis.

Let us start by obtaining the eikonal phase via a Fourier transform of our amplitudes.  
Then, following  Ref.~\cite{Bern:2020buy} we can solve Hamilton's equations for the impulse and color impulse and relate them  to derivatives of 
the eikonal phase.
The eikonal phase $\chi = \chi_1   
+ \chi_2 + \mathcal{O}(\alpha^3)$ is given by
\begin{align}
\chi_1 &= \frac{1}{4m_1m_2\sqrt{\sigma^2-1}}
\int \dd ^{2} \bm q \; e^{-i\bm{q}\cdot\bm{b}}\mathcal{A}^{\rm tree}(\bm{q}) \, , \nn \\
\chi_2 &= \frac{1}{4m_1m_2\sqrt{\sigma^2-1}} \int \dd ^{2} \bm q \; e^{-i\bm{q}\cdot\bm{b}}\mathcal{A}^{\bigtriangleup+\bigtriangledown}(\bm{q}) \, .
\end{align}
Using our amplitudes expressed in the center-of-mass frame (see Eq.~\eqref{fullTheoryM}) we find
\begin{align}
\chi_{1} &=- \frac{\xi E \alpha}{|\bm p|}
\am_1 \left(\ln \frac{\bm b^2}{L^2} \right)\mathcal{C}\left(\tree\right)
\,,  \\
\chi_{2}&= \frac{\pi  \xi E \alpha^2}{|\bm p|}
\,\frac{\am_2}{|\bm b|} \mathcal{C}\left(\boxy\right)\,,
\end{align} 
where in the first order eikonal phase we include a cutoff regulator $L$ as we did in Section \ref{sec:KMOC}. 
In the case without color, the integration is regulated 
via dimensional regularization, and the divergence is ignored,  because the derivative of the eikonal phase is always taken and they don't
contribute.  This is no longer the case here.  

We may now use the eikonal phase to obtain  classical observables. 
Generalizing the conjecture of Ref.~\cite{Bern:2020buy} to the color-charged case, the observables in question are the impulse
$\Delta \bm p$ and color impulse $\Delta c^a_i$, where
\begin{align}
\bm p(t=\infty) = \bm p + \Delta \bm p \,, 
\qquad
&\bm p(t=-\infty) = \bm p \,, \nn \\
c^a_i(t=\infty) = c^a_i + \Delta c^a_i \,, \qquad 
&c^a_i(t=-\infty) = c^a_i \,.
\end{align}
Inspired by the gravitational spinning case let us  decompose the impulse as 
\begin{equation}
\Delta \bm p = \Delta p_{\parallel} \frac{\bm p}{|\bm p|} + \Delta \bm p_\perp\,,
\end{equation}
where  $\Delta p_{\parallel}$ can be obtained from the on-shell condition $(\bm p + \Delta \bm p)^2 = \bm p^2$. Therefore,  
ignoring the mixing of spin  and  orbital angular momentum---which is absent in our case since the particle is 
spinless---the impulse and color impulse through $\mathcal{O}(\alpha^2)$ satisfy

\begin{align}
\Delta \bm p_\perp   &= -\{ \bm p_\perp, \chi \}
-\frac{1}{2}\,\{\chi, \{\bm p_\perp, \chi\} \} \,, \nn \\
\Delta c^a_1 &= -\{ c^a_1, \chi \}
-\frac{1}{2}\,\{\chi, \{c^a_1, \chi\} \} \,,
\label{eq:eikonalBrackets-2}
\end{align}
where in Eq.~\eqref{eq:eikonalBrackets-2} we use the definitions
\begin{align}
\{\bm p_\perp, g \} \equiv -\frac{\partial g}{\partial \bm b}
\,, \hskip 1 cm 
\{c_1^a,g \}  \equiv \,  
f^{abc}
\frac{\partial g}{\partial c^b_1} c_1^c  \, .
\label{Brackets}
\end{align}
The second term in the linear impulse doesn't contribute because the tree color structure commutes with itself but  we leave it there 
to keep the suggestive structure. It is then straightforward to show that the linear impulse will be reproduced here, the same way 
it was for the spinless QED case, simply by taking a replacement of electric for color charges.  We have compared both the impulse and the
color impulse, to the solution of the EOM,  and the integrated result of the NLO color impulse finding full agreement. 

Our calculation extends
the conjecture of Ref.~\cite{Bern:2020buy} to the domain of color.  
We may note that in this setting the momentum and the color are separately conserved. This is unlike the case for spinning particles, where only the sum $\bm J=\bm L+\bm S$ is conserved.   Due to the mixing of spin and orbital angular momentum,  it was possible to define the object $\mathcal{D}_{SL}\left(f, g \right) \equiv 
 - \, \bm S_1\, \cdot \left(\frac{\partial f}{\partial \bm S_1}\, \times\frac{\partial g}{\partial \bm L_b} \right) \,$ (where 
 $\bm S_1$ is the spin vector and $\bm L_{\bm b} \equiv \bm b \times \bm p$). Such an object was necessary to add terms of the form  $\mathcal{D}_{SL}\left(\chi, \{\bm o, \chi\} \right)$ and
$\{\bm o,\mathcal{D}_{SL}\left( \chi,  \chi \right) \}$.
In consequence,  the form of Eq.~(\ref{eq:eikonalBrackets-2}) is indeed simpler than its spin counterpart.

\section{Conclusions}
\label{sec:conclusions}
In this paper we have used the KMOC formalism and a matching procedure with a non-relativistic EFT to evaluate classical Yang-Mills 
observables. Using these 
approaches we have found that the eikonal phase conjecture of Ref.~\cite{Bern:2020buy} to the case of color is realized 
at NLO. On the KMOC side we have used the integrands already computed in Ref.~\cite{delaCruz:2020bbn} and performed a direct 
integration, while on the EFT side we have used unitarity adapting the formalism by Cheung-Rothstein-Solon 
\cite{Cheung:2018wkq} to the case of color charges.

The integration of the color and momentum impulses follows from a simple integral reduction and 
techniques successfully applied in gravity, e.g., in Ref.~\cite{Kalin:2020mvi}.  We have found that, as in the case of gravity, the integrals related only with the box and crossed
box vanish. However those related with the cut box contribute as expected. In order to expose the exponentiation of
the NLO color impulse we have used a cut-off regulator as in Ref.~\cite{delaCruz:2020bbn} to evaluate cut-box integrals.

Once the color decomposition has been performed and the classical relevant parts identified the matching procedure follows essentially the QED case. The Hamiltonian thus constructed was used to solve 
the  equations of motion and obtain the classical linear impulse and color impulse by direct integration. 
The results were in complete agreement to the evaluation using KMOC integrands. Finally, the eikonal phase construction 
matches the result of the KMOC and of EOM in  a rather elegant way giving more evidence of the observation  Ref.~\cite{Bern:2020buy}
that all physical observables are obtainable via simple manipulations of the scattering amplitude.

For the case of impulses it is also worth mentioning that the intricacies due to the mixing of color and kinematics in the KMOC
calculation are absent in the rather straightforward construction based on unitarity and EFT. However, for the  construction of the 
EFT it was crucial to employ coherent states to obtain the classical limit, so this aspect is common to both approaches as is the 
use  of the Lie algebra of the rescaled color factors. Obtaining higher order corrections in the KMOC formalism for Yang-Mills observables 
would be perhaps more efficient using  unitarity from the beginning as done in Refs.~\cite{Herrmann:2021lqe,Herrmann:2021tct} (for the gravitational case), benefiting from advances in relativistic integration.

Our results provide evidence in favor of the eikonal phase conjecture of Ref.~\cite{Bern:2020buy}, and so they call for the calculation of the 2-loop 
color impulse as a toy example towards the gravitational spin. Besides being  a toy model for gravitational dynamics, the classical
limit of Yang-Mills theory is useful to describe non-equilibrium plasma through kinetic theory, where color is treated as a continuous 
classical variable. In Ref.~\cite{delaCruz:2020cpc} solutions of kinetic equations were interpreted as classical limits of certain 
off-shell currents so it would be interesting to explore a Hamiltonian perspective to this problem.

\subsection*{ Acknowledgments:}
We thank Dimitrios Kosmopoulos for discussions and for sharing code regarding the solution of equations of motion. 
AL and TS are supported by the U.S. Department of Energy (DOE) under award number DE-SC0009937, and by the Mani L. Bhaumik Institute for 
Theoretical Physics. We thank Donal O'Connell and Radu Roiban for discussions and comments on the manuscript.   LDLC acknowledges financial support from the Open Physics Hub 
at the Physics and Astronomy Department in Bologna. 
We would like to thank the Galileo Galilei Institute for Theoretical Physics for hosting the workshop on Gravitational scattering, inspiral, and radiation where this collaboration was started.
Some of our figures were produced with the help of TikZ-Feynman~\cite{Ellis:2016jkw}.

\bibliographystyle{JHEP}

 \renewcommand\bibname{References} 
\ifdefined\phantomsection		
  \phantomsection  
\else
\fi
\addcontentsline{toc}{section}{References}

\bibliography{YMHam.bib}

\providecommand{\href}[2]{#2}\begingroup\raggedright\begin{thebibliography}{10}

\bibitem{Kosower:2018adc}
D.~A. Kosower, B.~Maybee and D.~O'Connell, \emph{{Amplitudes, Observables, and
  Classical Scattering}},
  \href{https://doi.org/10.1007/JHEP02(2019)137}{\emph{JHEP} {\bfseries 02}
  (2019) 137} [\href{https://arxiv.org/abs/1811.10950}{{\ttfamily
  1811.10950}}].

\bibitem{Maybee:2019jus}
B.~Maybee, D.~O'Connell and J.~Vines, \emph{{Observables and amplitudes for
  spinning particles and black holes}},
  \href{https://doi.org/10.1007/JHEP12(2019)156}{\emph{JHEP} {\bfseries 12}
  (2019) 156} [\href{https://arxiv.org/abs/1906.09260}{{\ttfamily
  1906.09260}}].

\bibitem{delaCruz:2020bbn}
L.~de~la Cruz, B.~Maybee, D.~O'Connell and A.~Ross, \emph{{Classical Yang-Mills
  observables from amplitudes}},
  \href{https://doi.org/10.1007/JHEP12(2020)076}{\emph{JHEP} {\bfseries 12}
  (2020) 076} [\href{https://arxiv.org/abs/2009.03842}{{\ttfamily
  2009.03842}}].

\bibitem{Cristofoli:2021vyo}
A.~Cristofoli, R.~Gonzo, D.~A. Kosower and D.~O'Connell, \emph{{Waveforms from
  Amplitudes}},  \href{https://arxiv.org/abs/2107.10193}{{\ttfamily
  2107.10193}}.

\bibitem{Manu:2020zxl}
A.~Manu, D.~Ghosh, A.~Laddha and P.~V. Athira, \emph{{Soft radiation from
  scattering amplitudes revisited}},
  \href{https://doi.org/10.1007/JHEP05(2021)056}{\emph{JHEP} {\bfseries 05}
  (2021) 056} [\href{https://arxiv.org/abs/2007.02077}{{\ttfamily
  2007.02077}}].

\bibitem{Herrmann:2021lqe}
E.~Herrmann, J.~Parra-Martinez, M.~S. Ruf and M.~Zeng, \emph{{Gravitational
  Bremsstrahlung from Reverse Unitarity}},
  \href{https://doi.org/10.1103/PhysRevLett.126.201602}{\emph{Phys. Rev. Lett.}
  {\bfseries 126} (2021) 201602}
  [\href{https://arxiv.org/abs/2101.07255}{{\ttfamily 2101.07255}}].

\bibitem{Herrmann:2021tct}
E.~Herrmann, J.~Parra-Martinez, M.~S. Ruf and M.~Zeng, \emph{{Radiative
  Classical Gravitational Observables at $\mathcal O(G^3)$ from Scattering
  Amplitudes}},  \href{https://arxiv.org/abs/2104.03957}{{\ttfamily
  2104.03957}}.

\bibitem{delaCruz:2020cpc}
L.~de~la Cruz, \emph{{Scattering amplitudes approach to hard thermal loops}},
  \href{https://doi.org/10.1103/PhysRevD.104.014013}{\emph{Phys. Rev. D}
  {\bfseries 104} (2021) 014013}
  [\href{https://arxiv.org/abs/2012.07714}{{\ttfamily 2012.07714}}].

\bibitem{Neill:2013wsa}
D.~Neill and I.~Z. Rothstein, \emph{{Classical Space-Times from the S Matrix}},
  \href{https://doi.org/10.1016/j.nuclphysb.2013.09.007}{\emph{Nucl. Phys. B}
  {\bfseries 877} (2013) 177}
  [\href{https://arxiv.org/abs/1304.7263}{{\ttfamily 1304.7263}}].

\bibitem{Cheung:2018wkq}
C.~Cheung, I.~Z. Rothstein and M.~P. Solon, \emph{{From Scattering Amplitudes
  to Classical Potentials in the Post-Minkowskian Expansion}},
  \href{https://doi.org/10.1103/PhysRevLett.121.251101}{\emph{Phys. Rev. Lett.}
  {\bfseries 121} (2018) 251101}
  [\href{https://arxiv.org/abs/1808.02489}{{\ttfamily 1808.02489}}].

\bibitem{Westpfahl:1985tsl}
K.~Westpfahl, \emph{{High-Speed Scattering of Charged and Uncharged Particles
  in General Relativity}},
  \href{https://doi.org/10.1002/prop.2190330802}{\emph{Fortsch. Phys.}
  {\bfseries 33} (1985) 417}.

\bibitem{Bern:2019nnu}
Z.~Bern, C.~Cheung, R.~Roiban, C.-H. Shen, M.~P. Solon and M.~Zeng,
  \emph{{Scattering Amplitudes and the Conservative Hamiltonian for Binary
  Systems at Third Post-Minkowskian Order}},
  \href{https://doi.org/10.1103/PhysRevLett.122.201603}{\emph{Phys. Rev. Lett.}
  {\bfseries 122} (2019) 201603}
  [\href{https://arxiv.org/abs/1901.04424}{{\ttfamily 1901.04424}}].

\bibitem{Bern:2019crd}
Z.~Bern, C.~Cheung, R.~Roiban, C.-H. Shen, M.~P. Solon and M.~Zeng,
  \emph{{Black Hole Binary Dynamics from the Double Copy and Effective
  Theory}}, \href{https://doi.org/10.1007/JHEP10(2019)206}{\emph{JHEP}
  {\bfseries 10} (2019) 206}
  [\href{https://arxiv.org/abs/1908.01493}{{\ttfamily 1908.01493}}].

\bibitem{Cheung:2020gyp}
C.~Cheung and M.~P. Solon, \emph{{Classical gravitational scattering at $
  \mathcal{O} $(G$^{3}$) from Feynman diagrams}},
  \href{https://doi.org/10.1007/JHEP06(2020)144}{\emph{JHEP} {\bfseries 06}
  (2020) 144} [\href{https://arxiv.org/abs/2003.08351}{{\ttfamily
  2003.08351}}].

\bibitem{Cristofoli:2019neg}
A.~Cristofoli, N.~E.~J. Bjerrum-Bohr, P.~H. Damgaard and P.~Vanhove,
  \emph{{Post-Minkowskian Hamiltonians in general relativity}},
  \href{https://doi.org/10.1103/PhysRevD.100.084040}{\emph{Phys. Rev. D}
  {\bfseries 100} (2019) 084040}
  [\href{https://arxiv.org/abs/1906.01579}{{\ttfamily 1906.01579}}].

\bibitem{Bjerrum-Bohr:2019kec}
N.~E.~J. Bjerrum-Bohr, A.~Cristofoli and P.~H. Damgaard,
  \emph{{Post-Minkowskian Scattering Angle in Einstein Gravity}},
  \href{https://doi.org/10.1007/JHEP08(2020)038}{\emph{JHEP} {\bfseries 08}
  (2020) 038} [\href{https://arxiv.org/abs/1910.09366}{{\ttfamily
  1910.09366}}].

\bibitem{Damgaard:2019lfh}
P.~H. Damgaard, K.~Haddad and A.~Helset, \emph{{Heavy Black Hole Effective
  Theory}}, \href{https://doi.org/10.1007/JHEP11(2019)070}{\emph{JHEP}
  {\bfseries 11} (2019) 070}
  [\href{https://arxiv.org/abs/1908.10308}{{\ttfamily 1908.10308}}].

\bibitem{Aoude:2020onz}
R.~Aoude, K.~Haddad and A.~Helset, \emph{{On-shell heavy particle effective
  theories}}, \href{https://doi.org/10.1007/JHEP05(2020)051}{\emph{JHEP}
  {\bfseries 05} (2020) 051}
  [\href{https://arxiv.org/abs/2001.09164}{{\ttfamily 2001.09164}}].

\bibitem{Haddad:2020tvs}
K.~Haddad and A.~Helset, \emph{{The double copy for heavy particles}},
  \href{https://doi.org/10.1103/PhysRevLett.125.181603}{\emph{Phys. Rev. Lett.}
  {\bfseries 125} (2020) 181603}
  [\href{https://arxiv.org/abs/2005.13897}{{\ttfamily 2005.13897}}].

\bibitem{Kalin:2019rwq}
G.~K\"alin and R.~A. Porto, \emph{{From Boundary Data to Bound States}},
  \href{https://doi.org/10.1007/JHEP01(2020)072}{\emph{JHEP} {\bfseries 01}
  (2020) 072} [\href{https://arxiv.org/abs/1910.03008}{{\ttfamily
  1910.03008}}].

\bibitem{Kalin:2019inp}
G.~K\"alin and R.~A. Porto, \emph{{From boundary data to bound states. Part II.
  Scattering angle to dynamical invariants (with twist)}},
  \href{https://doi.org/10.1007/JHEP02(2020)120}{\emph{JHEP} {\bfseries 02}
  (2020) 120} [\href{https://arxiv.org/abs/1911.09130}{{\ttfamily
  1911.09130}}].

\bibitem{Kalin:2020mvi}
G.~K\"alin and R.~A. Porto, \emph{{Post-Minkowskian Effective Field Theory for
  Conservative Binary Dynamics}},
  \href{https://doi.org/10.1007/JHEP11(2020)106}{\emph{JHEP} {\bfseries 11}
  (2020) 106} [\href{https://arxiv.org/abs/2006.01184}{{\ttfamily
  2006.01184}}].

\bibitem{Kalin:2020fhe}
G.~K\"alin, Z.~Liu and R.~A. Porto, \emph{{Conservative Dynamics of Binary
  Systems to Third Post-Minkowskian Order from the Effective Field Theory
  Approach}}, \href{https://doi.org/10.1103/PhysRevLett.125.261103}{\emph{Phys.
  Rev. Lett.} {\bfseries 125} (2020) 261103}
  [\href{https://arxiv.org/abs/2007.04977}{{\ttfamily 2007.04977}}].

\bibitem{Dlapa:2021npj}
C.~Dlapa, G.~K\"alin, Z.~Liu and R.~A. Porto, \emph{{Dynamics of Binary Systems
  to Fourth Post-Minkowskian Order from the Effective Field Theory Approach}},
  \href{https://arxiv.org/abs/2106.08276}{{\ttfamily 2106.08276}}.

\bibitem{Mogull:2020sak}
G.~Mogull, J.~Plefka and J.~Steinhoff, \emph{{Classical black hole scattering
  from a worldline quantum field theory}},
  \href{https://doi.org/10.1007/JHEP02(2021)048}{\emph{JHEP} {\bfseries 02}
  (2021) 048} [\href{https://arxiv.org/abs/2010.02865}{{\ttfamily
  2010.02865}}].

\bibitem{Bern:2021dqo}
Z.~Bern, J.~Parra-Martinez, R.~Roiban, M.~S. Ruf, C.-H. Shen, M.~P. Solon
  et~al., \emph{{Scattering Amplitudes and Conservative Binary Dynamics at
  ${\cal O}(G^4)$}},
  \href{https://doi.org/10.1103/PhysRevLett.126.171601}{\emph{Phys. Rev. Lett.}
  {\bfseries 126} (2021) 171601}
  [\href{https://arxiv.org/abs/2101.07254}{{\ttfamily 2101.07254}}].

\bibitem{Caron-Huot:2018ape}
S.~Caron-Huot and Z.~Zahraee, \emph{{Integrability of Black Hole Orbits in
  Maximal Supergravity}},
  \href{https://doi.org/10.1007/JHEP07(2019)179}{\emph{JHEP} {\bfseries 07}
  (2019) 179} [\href{https://arxiv.org/abs/1810.04694}{{\ttfamily
  1810.04694}}].

\bibitem{Bern:2020gjj}
Z.~Bern, H.~Ita, J.~Parra-Martinez and M.~S. Ruf, \emph{{Universality in the
  classical limit of massless gravitational scattering}},
  \href{https://doi.org/10.1103/PhysRevLett.125.031601}{\emph{Phys. Rev. Lett.}
  {\bfseries 125} (2020) 031601}
  [\href{https://arxiv.org/abs/2002.02459}{{\ttfamily 2002.02459}}].

\bibitem{Parra-Martinez:2020dzs}
J.~Parra-Martinez, M.~S. Ruf and M.~Zeng, \emph{{Extremal black hole scattering
  at $\mathcal{O}(G^3)$: graviton dominance, eikonal exponentiation, and
  differential equations}},
  \href{https://doi.org/10.1007/JHEP11(2020)023}{\emph{JHEP} {\bfseries 11}
  (2020) 023} [\href{https://arxiv.org/abs/2005.04236}{{\ttfamily
  2005.04236}}].

\bibitem{Carrillo-Gonzalez:2021mqj}
M.~Carrillo-Gonz\'alez, C.~de~Rham and A.~J. Tolley, \emph{{Scattering
  Amplitudes for Binary Systems beyond GR}},
  \href{https://arxiv.org/abs/2107.11384}{{\ttfamily 2107.11384}}.

\bibitem{Gonzalez:2020krh}
M.~C. Gonzalez, Q.~Liang and M.~Trodden, \emph{{An Effective Field Theory for
  Binary Cosmic Strings}},  \href{https://arxiv.org/abs/2010.15913}{{\ttfamily
  2010.15913}}.

\bibitem{Loebbert:2020aos}
F.~Loebbert, J.~Plefka, C.~Shi and T.~Wang, \emph{{Three-Body Effective
  Potential in General Relativity at 2PM and Resulting PN Contributions}},
  \href{https://arxiv.org/abs/2012.14224}{{\ttfamily 2012.14224}}.

\bibitem{DiVecchia:2020ymx}
P.~Di~Vecchia, C.~Heissenberg, R.~Russo and G.~Veneziano, \emph{{Universality
  of ultra-relativistic gravitational scattering}},
  \href{https://doi.org/10.1016/j.physletb.2020.135924}{\emph{Phys. Lett. B}
  {\bfseries 811} (2020) 135924}
  [\href{https://arxiv.org/abs/2008.12743}{{\ttfamily 2008.12743}}].

\bibitem{DiVecchia:2021ndb}
P.~Di~Vecchia, C.~Heissenberg, R.~Russo and G.~Veneziano, \emph{{Radiation
  Reaction from Soft Theorems}},
  \href{https://arxiv.org/abs/2101.05772}{{\ttfamily 2101.05772}}.

\bibitem{Damour:2020tta}
T.~Damour, \emph{{Radiative contribution to classical gravitational scattering
  at the third order in $G$}},
  \href{https://doi.org/10.1103/PhysRevD.102.124008}{\emph{Phys. Rev. D}
  {\bfseries 102} (2020) 124008}
  [\href{https://arxiv.org/abs/2010.01641}{{\ttfamily 2010.01641}}].

\bibitem{Jakobsen:2021smu}
G.~U. Jakobsen, G.~Mogull, J.~Plefka and J.~Steinhoff, \emph{{Classical
  Gravitational Bremsstrahlung from a Worldline Quantum Field Theory}},
  \href{https://doi.org/10.1103/PhysRevLett.126.201103}{\emph{Phys. Rev. Lett.}
  {\bfseries 126} (2021) 201103}
  [\href{https://arxiv.org/abs/2101.12688}{{\ttfamily 2101.12688}}].

\bibitem{Mougiakakos:2021ckm}
S.~Mougiakakos, M.~M. Riva and F.~Vernizzi, \emph{{Gravitational Bremsstrahlung
  in the post-Minkowskian effective field theory}},
  \href{https://doi.org/10.1103/PhysRevD.104.024041}{\emph{Phys. Rev. D}
  {\bfseries 104} (2021) 024041}
  [\href{https://arxiv.org/abs/2102.08339}{{\ttfamily 2102.08339}}].

\bibitem{Bjerrum-Bohr:2021din}
N.~E.~J. Bjerrum-Bohr, P.~H. Damgaard, L.~Plant\'e and P.~Vanhove, \emph{{The
  Amplitude for Classical Gravitational Scattering at Third Post-Minkowskian
  Order}},  \href{https://arxiv.org/abs/2105.05218}{{\ttfamily 2105.05218}}.

\bibitem{Haddad:2020que}
K.~Haddad and A.~Helset, \emph{{Tidal effects in quantum field theory}},
  \href{https://doi.org/10.1007/JHEP12(2020)024}{\emph{JHEP} {\bfseries 12}
  (2020) 024} [\href{https://arxiv.org/abs/2008.04920}{{\ttfamily
  2008.04920}}].

\bibitem{Aoude:2020ygw}
R.~Aoude, K.~Haddad and A.~Helset, \emph{{Tidal effects for spinning
  particles}}, \href{https://doi.org/10.1007/JHEP03(2021)097}{\emph{JHEP}
  {\bfseries 03} (2021) 097}
  [\href{https://arxiv.org/abs/2012.05256}{{\ttfamily 2012.05256}}].

\bibitem{AccettulliHuber:2020dal}
M.~Accettulli~Huber, A.~Brandhuber, S.~De~Angelis and G.~Travaglini,
  \emph{{From amplitudes to gravitational radiation with cubic interactions and
  tidal effects}},
  \href{https://doi.org/10.1103/PhysRevD.103.045015}{\emph{Phys. Rev. D}
  {\bfseries 103} (2021) 045015}
  [\href{https://arxiv.org/abs/2012.06548}{{\ttfamily 2012.06548}}].

\bibitem{Kalin:2020lmz}
G.~K\"alin, Z.~Liu and R.~A. Porto, \emph{{Conservative Tidal Effects in
  Compact Binary Systems to Next-to-Leading Post-Minkowskian Order}},
  \href{https://doi.org/10.1103/PhysRevD.102.124025}{\emph{Phys. Rev. D}
  {\bfseries 102} (2020) 124025}
  [\href{https://arxiv.org/abs/2008.06047}{{\ttfamily 2008.06047}}].

\bibitem{Cheung:2020sdj}
C.~Cheung and M.~P. Solon, \emph{{Tidal Effects in the Post-Minkowskian
  Expansion}},
  \href{https://doi.org/10.1103/PhysRevLett.125.191601}{\emph{Phys. Rev. Lett.}
  {\bfseries 125} (2020) 191601}
  [\href{https://arxiv.org/abs/2006.06665}{{\ttfamily 2006.06665}}].

\bibitem{Cheung:2020gbf}
C.~Cheung, N.~Shah and M.~P. Solon, \emph{{Mining the Geodesic Equation for
  Scattering Data}},
  \href{https://doi.org/10.1103/PhysRevD.103.024030}{\emph{Phys. Rev. D}
  {\bfseries 103} (2021) 024030}
  [\href{https://arxiv.org/abs/2010.08568}{{\ttfamily 2010.08568}}].

\bibitem{Bern:2020uwk}
Z.~Bern, J.~Parra-Martinez, R.~Roiban, E.~Sawyer and C.-H. Shen, \emph{{Leading
  Nonlinear Tidal Effects and Scattering Amplitudes}},
  \href{https://doi.org/10.1007/JHEP05(2021)188}{\emph{JHEP} {\bfseries 05}
  (2021) 188} [\href{https://arxiv.org/abs/2010.08559}{{\ttfamily
  2010.08559}}].

\bibitem{Holstein:2008sx}
B.~R. Holstein and A.~Ross, \emph{{Spin Effects in Long Range Gravitational
  Scattering}},  \href{https://arxiv.org/abs/0802.0716}{{\ttfamily 0802.0716}}.

\bibitem{Vaidya:2014kza}
V.~Vaidya, \emph{{Gravitational spin Hamiltonians from the S matrix}},
  \href{https://doi.org/10.1103/PhysRevD.91.024017}{\emph{Phys. Rev. D}
  {\bfseries 91} (2015) 024017}
  [\href{https://arxiv.org/abs/1410.5348}{{\ttfamily 1410.5348}}].

\bibitem{Guevara:2017csg}
A.~Guevara, \emph{{Holomorphic Classical Limit for Spin Effects in
  Gravitational and Electromagnetic Scattering}},
  \href{https://doi.org/10.1007/JHEP04(2019)033}{\emph{JHEP} {\bfseries 04}
  (2019) 033} [\href{https://arxiv.org/abs/1706.02314}{{\ttfamily
  1706.02314}}].

\bibitem{Guevara:2018wpp}
A.~Guevara, A.~Ochirov and J.~Vines, \emph{{Scattering of Spinning Black Holes
  from Exponentiated Soft Factors}},
  \href{https://doi.org/10.1007/JHEP09(2019)056}{\emph{JHEP} {\bfseries 09}
  (2019) 056} [\href{https://arxiv.org/abs/1812.06895}{{\ttfamily
  1812.06895}}].

\bibitem{Chung:2018kqs}
M.-Z. Chung, Y.-T. Huang, J.-W. Kim and S.~Lee, \emph{{The simplest massive
  S-matrix: from minimal coupling to Black Holes}},
  \href{https://doi.org/10.1007/JHEP04(2019)156}{\emph{JHEP} {\bfseries 04}
  (2019) 156} [\href{https://arxiv.org/abs/1812.08752}{{\ttfamily
  1812.08752}}].

\bibitem{Chung:2019duq}
M.-Z. Chung, Y.-T. Huang and J.-W. Kim, \emph{{Classical potential for general
  spinning bodies}}, \href{https://doi.org/10.1007/JHEP09(2020)074}{\emph{JHEP}
  {\bfseries 09} (2020) 074}
  [\href{https://arxiv.org/abs/1908.08463}{{\ttfamily 1908.08463}}].

\bibitem{Chung:2020rrz}
M.-Z. Chung, Y.-t. Huang, J.-W. Kim and S.~Lee, \emph{{Complete Hamiltonian for
  spinning binary systems at first post-Minkowskian order}},
  \href{https://doi.org/10.1007/JHEP05(2020)105}{\emph{JHEP} {\bfseries 05}
  (2020) 105} [\href{https://arxiv.org/abs/2003.06600}{{\ttfamily
  2003.06600}}].

\bibitem{Guevara:2019fsj}
A.~Guevara, A.~Ochirov and J.~Vines, \emph{{Black-hole scattering with general
  spin directions from minimal-coupling amplitudes}},
  \href{https://doi.org/10.1103/PhysRevD.100.104024}{\emph{Phys. Rev. D}
  {\bfseries 100} (2019) 104024}
  [\href{https://arxiv.org/abs/1906.10071}{{\ttfamily 1906.10071}}].

\bibitem{Vines:2017hyw}
J.~Vines, \emph{{Scattering of two spinning black holes in post-Minkowskian
  gravity, to all orders in spin, and effective-one-body mappings}},
  \href{https://doi.org/10.1088/1361-6382/aaa3a8}{\emph{Class. Quant. Grav.}
  {\bfseries 35} (2018) 084002}
  [\href{https://arxiv.org/abs/1709.06016}{{\ttfamily 1709.06016}}].

\bibitem{Bern:2020buy}
Z.~Bern, A.~Luna, R.~Roiban, C.-H. Shen and M.~Zeng, \emph{{Spinning Black Hole
  Binary Dynamics, Scattering Amplitudes and Effective Field Theory}},
  \href{https://arxiv.org/abs/2005.03071}{{\ttfamily 2005.03071}}.

\bibitem{Liu:2021zxr}
Z.~Liu, R.~A. Porto and Z.~Yang, \emph{{Spin Effects in the Effective Field
  Theory Approach to Post-Minkowskian Conservative Dynamics}},
  \href{https://doi.org/10.1007/JHEP06(2021)012}{\emph{JHEP} {\bfseries 06}
  (2021) 012} [\href{https://arxiv.org/abs/2102.10059}{{\ttfamily
  2102.10059}}].

\bibitem{Kosmopoulos:2021zoq}
D.~Kosmopoulos and A.~Luna, \emph{{Quadratic-in-spin Hamiltonian at $
  \mathcal{O} $(G$^{2}$) from scattering amplitudes}},
  \href{https://doi.org/10.1007/JHEP07(2021)037}{\emph{JHEP} {\bfseries 07}
  (2021) 037} [\href{https://arxiv.org/abs/2102.10137}{{\ttfamily
  2102.10137}}].

\bibitem{Jakobsen:2021lvp}
G.~U. Jakobsen, G.~Mogull, J.~Plefka and J.~Steinhoff, \emph{{Gravitational
  Bremsstrahlung and Hidden Supersymmetry of Spinning Bodies}},
  \href{https://arxiv.org/abs/2106.10256}{{\ttfamily 2106.10256}}.

\bibitem{Chen:2021huj}
B.-T. Chen, M.-Z. Chung, Y.-t. Huang and M.~K. Tam, \emph{{Minimal spin
  deflection of Kerr-Newman and Supersymmetric black hole}},
  \href{https://arxiv.org/abs/2106.12518}{{\ttfamily 2106.12518}}.

\bibitem{Bautista:2021wfy}
Y.~F. Bautista, A.~Guevara, C.~Kavanagh and J.~Vines, \emph{{From Scattering in
  Black Hole Backgrounds to Higher-Spin Amplitudes: Part I}},
  \href{https://arxiv.org/abs/2107.10179}{{\ttfamily 2107.10179}}.

\bibitem{Aoude:2021oqj}
R.~Aoude and A.~Ochirov, \emph{{Classical observables from coherent-spin
  amplitudes}},  \href{https://arxiv.org/abs/2108.01649}{{\ttfamily
  2108.01649}}.

\bibitem{Chiodaroli:2021eug}
M.~Chiodaroli, H.~Johansson and P.~Pichini, \emph{{Compton Black-Hole
  Scattering for $s \leq 5/2$}},
  \href{https://arxiv.org/abs/2107.14779}{{\ttfamily 2107.14779}}.

\bibitem{Amati:1990xe}
D.~Amati, M.~Ciafaloni and G.~Veneziano, \emph{{Higher Order Gravitational
  Deflection and Soft Bremsstrahlung in Planckian Energy Superstring
  Collisions}}, \href{https://doi.org/10.1016/0550-3213(90)90375-N}{\emph{Nucl.
  Phys. B} {\bfseries 347} (1990) 550}.

\bibitem{Melville:2013qca}
S.~Melville, S.~G. Naculich, H.~J. Schnitzer and C.~D. White, \emph{{Wilson
  line approach to gravity in the high energy limit}},
  \href{https://doi.org/10.1103/PhysRevD.89.025009}{\emph{Phys. Rev. D}
  {\bfseries 89} (2014) 025009}
  [\href{https://arxiv.org/abs/1306.6019}{{\ttfamily 1306.6019}}].

\bibitem{Luna:2016idw}
A.~Luna, S.~Melville, S.~G. Naculich and C.~D. White, \emph{{Next-to-soft
  corrections to high energy scattering in QCD and gravity}},
  \href{https://doi.org/10.1007/JHEP01(2017)052}{\emph{JHEP} {\bfseries 01}
  (2017) 052} [\href{https://arxiv.org/abs/1611.02172}{{\ttfamily
  1611.02172}}].

\bibitem{Akhoury:2013yua}
R.~Akhoury, R.~Saotome and G.~Sterman, \emph{{High Energy Scattering in
  Perturbative Quantum Gravity at Next to Leading Power}},
  \href{https://doi.org/10.1103/PhysRevD.103.064036}{\emph{Phys. Rev. D}
  {\bfseries 103} (2021) 064036}
  [\href{https://arxiv.org/abs/1308.5204}{{\ttfamily 1308.5204}}].

\bibitem{KoemansCollado:2019ggb}
A.~Koemans~Collado, P.~Di~Vecchia and R.~Russo, \emph{{Revisiting the second
  post-Minkowskian eikonal and the dynamics of binary black holes}},
  \href{https://doi.org/10.1103/PhysRevD.100.066028}{\emph{Phys. Rev. D}
  {\bfseries 100} (2019) 066028}
  [\href{https://arxiv.org/abs/1904.02667}{{\ttfamily 1904.02667}}].

\bibitem{Cristofoli:2020uzm}
A.~Cristofoli, P.~H. Damgaard, P.~Di~Vecchia and C.~Heissenberg,
  \emph{{Second-order Post-Minkowskian scattering in arbitrary dimensions}},
  \href{https://doi.org/10.1007/JHEP07(2020)122}{\emph{JHEP} {\bfseries 07}
  (2020) 122} [\href{https://arxiv.org/abs/2003.10274}{{\ttfamily
  2003.10274}}].

\bibitem{DiVecchia:2019myk}
P.~Di~Vecchia, A.~Luna, S.~G. Naculich, R.~Russo, G.~Veneziano and C.~D. White,
  \emph{{A tale of two exponentiations in ${\cal N}=8$ supergravity}},
  \href{https://doi.org/10.1016/j.physletb.2019.134927}{\emph{Phys. Lett. B}
  {\bfseries 798} (2019) 134927}
  [\href{https://arxiv.org/abs/1908.05603}{{\ttfamily 1908.05603}}].

\bibitem{DiVecchia:2019kta}
P.~Di~Vecchia, S.~G. Naculich, R.~Russo, G.~Veneziano and C.~D. White, \emph{{A
  tale of two exponentiations in $ \mathcal{N} $ = 8 supergravity at subleading
  level}}, \href{https://doi.org/10.1007/JHEP03(2020)173}{\emph{JHEP}
  {\bfseries 03} (2020) 173}
  [\href{https://arxiv.org/abs/1911.11716}{{\ttfamily 1911.11716}}].

\bibitem{DiVecchia:2021bdo}
P.~Di~Vecchia, C.~Heissenberg, R.~Russo and G.~Veneziano, \emph{{The Eikonal
  Approach to Gravitational Scattering and Radiation at $\mathcal O(G^3)$}},
  \href{https://arxiv.org/abs/2104.03256}{{\ttfamily 2104.03256}}.

\bibitem{Heissenberg:2021tzo}
C.~Heissenberg, \emph{{Infrared Divergences and the Eikonal}},
  \href{https://arxiv.org/abs/2105.04594}{{\ttfamily 2105.04594}}.

\bibitem{Damgaard:2021ipf}
P.~H. Damgaard, L.~Plante and P.~Vanhove, \emph{{On an Exponential
  Representation of the Gravitational S-Matrix}},
  \href{https://arxiv.org/abs/2107.12891}{{\ttfamily 2107.12891}}.

\bibitem{Vines:2018gqi}
J.~Vines, J.~Steinhoff and A.~Buonanno, \emph{{Spinning-black-hole scattering
  and the test-black-hole limit at second post-Minkowskian order}},
  \href{https://doi.org/10.1103/PhysRevD.99.064054}{\emph{Phys. Rev. D}
  {\bfseries 99} (2019) 064054}
  [\href{https://arxiv.org/abs/1812.00956}{{\ttfamily 1812.00956}}].

\bibitem{Siemonsen:2019dsu}
N.~Siemonsen and J.~Vines, \emph{{Test black holes, scattering amplitudes and
  perturbations of Kerr spacetime}},
  \href{https://doi.org/10.1103/PhysRevD.101.064066}{\emph{Phys. Rev. D}
  {\bfseries 101} (2020) 064066}
  [\href{https://arxiv.org/abs/1909.07361}{{\ttfamily 1909.07361}}].

\bibitem{Bern:2019prr}
Z.~Bern, J.~J. Carrasco, M.~Chiodaroli, H.~Johansson and R.~Roiban, \emph{{The
  Duality Between Color and Kinematics and its Applications}},
  \href{https://arxiv.org/abs/1909.01358}{{\ttfamily 1909.01358}}.

\bibitem{Goldberger:2016iau}
W.~D. Goldberger and A.~K. Ridgway, \emph{{Radiation and the classical double
  copy for color charges}},
  \href{https://doi.org/10.1103/PhysRevD.95.125010}{\emph{Phys. Rev. D}
  {\bfseries 95} (2017) 125010}
  [\href{https://arxiv.org/abs/1611.03493}{{\ttfamily 1611.03493}}].

\bibitem{Goldberger:2017frp}
W.~D. Goldberger, S.~G. Prabhu and J.~O. Thompson, \emph{{Classical gluon and
  graviton radiation from the bi-adjoint scalar double copy}},
  \href{https://doi.org/10.1103/PhysRevD.96.065009}{\emph{Phys. Rev. D}
  {\bfseries 96} (2017) 065009}
  [\href{https://arxiv.org/abs/1705.09263}{{\ttfamily 1705.09263}}].

\bibitem{Goldberger:2017vcg}
W.~D. Goldberger and A.~K. Ridgway, \emph{{Bound states and the classical
  double copy}}, \href{https://doi.org/10.1103/PhysRevD.97.085019}{\emph{Phys.
  Rev. D} {\bfseries 97} (2018) 085019}
  [\href{https://arxiv.org/abs/1711.09493}{{\ttfamily 1711.09493}}].

\bibitem{Goldberger:2017ogt}
W.~D. Goldberger, J.~Li and S.~G. Prabhu, \emph{{Spinning particles, axion
  radiation, and the classical double copy}},
  \href{https://doi.org/10.1103/PhysRevD.97.105018}{\emph{Phys. Rev. D}
  {\bfseries 97} (2018) 105018}
  [\href{https://arxiv.org/abs/1712.09250}{{\ttfamily 1712.09250}}].

\bibitem{Chester:2017vcz}
D.~Chester, \emph{{Radiative double copy for Einstein-Yang-Mills theory}},
  \href{https://doi.org/10.1103/PhysRevD.97.084025}{\emph{Phys. Rev. D}
  {\bfseries 97} (2018) 084025}
  [\href{https://arxiv.org/abs/1712.08684}{{\ttfamily 1712.08684}}].

\bibitem{Shen:2018ebu}
C.-H. Shen, \emph{{Gravitational Radiation from Color-Kinematics Duality}},
  \href{https://doi.org/10.1007/JHEP11(2018)162}{\emph{JHEP} {\bfseries 11}
  (2018) 162} [\href{https://arxiv.org/abs/1806.07388}{{\ttfamily
  1806.07388}}].

\bibitem{Luna:2016hge}
A.~Luna, R.~Monteiro, I.~Nicholson, A.~Ochirov, D.~O'Connell, N.~Westerberg
  et~al., \emph{{Perturbative spacetimes from Yang-Mills theory}},
  \href{https://doi.org/10.1007/JHEP04(2017)069}{\emph{JHEP} {\bfseries 04}
  (2017) 069} [\href{https://arxiv.org/abs/1611.07508}{{\ttfamily
  1611.07508}}].

\bibitem{Luna:2017dtq}
A.~Luna, I.~Nicholson, D.~O'Connell and C.~D. White, \emph{{Inelastic Black
  Hole Scattering from Charged Scalar Amplitudes}},
  \href{https://doi.org/10.1007/JHEP03(2018)044}{\emph{JHEP} {\bfseries 03}
  (2018) 044} [\href{https://arxiv.org/abs/1711.03901}{{\ttfamily
  1711.03901}}].

\bibitem{CarrilloGonzalez:2018ejf}
M.~Carrillo~Gonz\'alez, R.~Penco and M.~Trodden, \emph{{Radiation of scalar
  modes and the classical double copy}},
  \href{https://doi.org/10.1007/JHEP11(2018)065}{\emph{JHEP} {\bfseries 11}
  (2018) 065} [\href{https://arxiv.org/abs/1809.04611}{{\ttfamily
  1809.04611}}].

\bibitem{Plefka:2018dpa}
J.~Plefka, J.~Steinhoff and W.~Wormsbecher, \emph{{Effective action of dilaton
  gravity as the classical double copy of Yang-Mills theory}},
  \href{https://doi.org/10.1103/PhysRevD.99.024021}{\emph{Phys. Rev. D}
  {\bfseries 99} (2019) 024021}
  [\href{https://arxiv.org/abs/1807.09859}{{\ttfamily 1807.09859}}].

\bibitem{Plefka:2019hmz}
J.~Plefka, C.~Shi, J.~Steinhoff and T.~Wang, \emph{{Breakdown of the classical
  double copy for the effective action of dilaton-gravity at NNLO}},
  \href{https://doi.org/10.1103/PhysRevD.100.086006}{\emph{Phys. Rev. D}
  {\bfseries 100} (2019) 086006}
  [\href{https://arxiv.org/abs/1906.05875}{{\ttfamily 1906.05875}}].

\bibitem{Monteiro:2014cda}
R.~Monteiro, D.~O'Connell and C.~D. White, \emph{{Black holes and the double
  copy}}, \href{https://doi.org/10.1007/JHEP12(2014)056}{\emph{JHEP} {\bfseries
  12} (2014) 056} [\href{https://arxiv.org/abs/1410.0239}{{\ttfamily
  1410.0239}}].

\bibitem{Luna:2018dpt}
A.~Luna, R.~Monteiro, I.~Nicholson and D.~O'Connell, \emph{{Type D Spacetimes
  and the Weyl Double Copy}},
  \href{https://doi.org/10.1088/1361-6382/ab03e6}{\emph{Class. Quant. Grav.}
  {\bfseries 36} (2019) 065003}
  [\href{https://arxiv.org/abs/1810.08183}{{\ttfamily 1810.08183}}].

\bibitem{Keeler:2020rcv}
C.~Keeler, T.~Manton and N.~Monga, \emph{{From Navier-Stokes to Maxwell via
  Einstein}}, \href{https://doi.org/10.1007/JHEP08(2020)147}{\emph{JHEP}
  {\bfseries 08} (2020) 147}
  [\href{https://arxiv.org/abs/2005.04242}{{\ttfamily 2005.04242}}].

\bibitem{Cheung:2020djz}
C.~Cheung and J.~Mangan, \emph{{Scattering Amplitudes and the Navier-Stokes
  Equation}},  \href{https://arxiv.org/abs/2010.15970}{{\ttfamily 2010.15970}}.

\bibitem{Bjerrum-Bohr:2018xdl}
N.~E.~J. Bjerrum-Bohr, P.~H. Damgaard, G.~Festuccia, L.~Plant\'e and
  P.~Vanhove, \emph{{General Relativity from Scattering Amplitudes}},
  \href{https://doi.org/10.1103/PhysRevLett.121.171601}{\emph{Phys. Rev. Lett.}
  {\bfseries 121} (2018) 171601}
  [\href{https://arxiv.org/abs/1806.04920}{{\ttfamily 1806.04920}}].

\bibitem{Bjerrum-Bohr:2021vuf}
N.~E.~J. Bjerrum-Bohr, P.~H. Damgaard, L.~Plant\'e and P.~Vanhove,
  \emph{{Classical Gravity from Loop Amplitudes}},
  \href{https://arxiv.org/abs/2104.04510}{{\ttfamily 2104.04510}}.

\bibitem{Wong:1970fu}
S.~K. Wong, \emph{{Field and particle equations for the classical Yang-Mills
  field and particles with isotopic spin}},
  \href{https://doi.org/10.1007/BF02892134}{\emph{Nuovo Cim. A} {\bfseries 65}
  (1970) 689}.

\bibitem{Balachandran:1976ya}
A.~P. Balachandran, P.~Salomonson, B.-S. Skagerstam and J.-O. Winnberg,
  \emph{{Classical Description of Particle Interacting with Nonabelian Gauge
  Field}}, \href{https://doi.org/10.1103/PhysRevD.15.2308}{\emph{Phys. Rev. D}
  {\bfseries 15} (1977) 2308}.

\bibitem{Balachandran:1977ub}
A.~P. Balachandran, S.~Borchardt and A.~Stern, \emph{{Lagrangian and
  Hamiltonian Descriptions of Yang-Mills Particles}},
  \href{https://doi.org/10.1103/PhysRevD.17.3247}{\emph{Phys. Rev. D}
  {\bfseries 17} (1978) 3247}.

\bibitem{Ellis:2016jkw}
J.~Ellis, \emph{{TikZ-Feynman: Feynman diagrams with TikZ}},
  \href{https://doi.org/10.1016/j.cpc.2016.08.019}{\emph{Comput. Phys. Commun.}
  {\bfseries 210} (2017) 103}
  [\href{https://arxiv.org/abs/1601.05437}{{\ttfamily 1601.05437}}].

\end{thebibliography}\endgroup

\end{document}